\documentclass[useAMS,usenatbib,usegraphicx,A4paper]{mn2e}
\usepackage{times} 

\title[Evidence for episodic warm outflow from IRAS~08470-4321]{Evidence for episodic warm
  outflowing CO gas from the intermediate mass young stellar object
  IRAS~08470-4321\thanks{Based on observations collected at the Very
    Large Telescope of the European Southern Observatory at La Silla
    and Paranal, Chile (ESO Programme I64.I-0605 and I79.C-0151}}
\author[W.-F. Thi et al.]{W.-F. Thi$^{1}$ \thanks{E-mail:
    wfdt@roe.ac.uk}, E.~F.~van~Dishoeck$^{2,3}$,
  K.~M.~Pontoppidan$^{4}$, E.~Dartois$^{5}$ \\
  $^{1}$\thanks{Scottish Universities Physics Alliances} SUPA, Institute for Astronomy, Royal Observatory Edinburgh, University of Edinburgh, Blackford Hill, Edinburgh, EH9 3HJ, UK\\
  $^{2}$Leiden Observatory, Leiden University, P. O.  Box 9513, 2300, Leiden, The Netherlands \\
  $^{3}$Max Planck Institut f\"{u}r Extrarerrestrische Physics, Postfach 1312, 85741, Garching, Germany \\
  $^{4}$Division of Geological and Planetary Science, MS 150-21, California Institute of Technology, CA 91125, USA\\
  $^{5}$Astrochimie Exp\'erimentale, Institut d'Astrophysique
  Spatiale, Universit\'e Paris-Sud, B\^at. 121,
  F-91405 Orsay, France\\
}
\begin{document}

\date{Accepted 2009. Received 2010}

\pagerange{\pageref{firstpage}--\pageref{lastpage}} \pubyear{2009}

\maketitle

\label{firstpage}
\begin{abstract}
  We present a $R\simeq$10,000 $M$-band spectrum of LLN~19
  (IRAS~08470-4321), a heavily embedded intermediate-mass young
  stellar object located in the Vela Molecular Cloud, obtained with
  {\em VLT-ISAAC}. The data were fitted by a 2-slab cold-hot model and
  a wind model. The spectrum exhibits deep broad ro-vibrational
  absorption lines of $^{12}$CO $v=1\leftarrow 0$ and $^{13}$CO
  $v=1\leftarrow 0$. A weak CO ice feature at 4.67 $\mu$m is also
  detected. Differences in velocity indicate that the warm gas is
  distinct from the cold millimeter emitting gas, which may be
  associated with the absorption by cooler gas (45~K). The outflowing
  warm gas at 300--400~K and with a mass-loss rate varying between
  0.48 $\times$ 10$^{-7}$ and 4.2 $\times$ 10$^{-7}$ M$_\odot$
  yr$^{-1}$ can explain most of the absorption. Several absorption
  lines were spectrally resolved in subsequent spectra obtained with
  the {\em VLT-CRIRES} instrument. Multiple absorption substructures
  in the high-resolution ($R$=100,000) spectra indicate that the
  mass-loss is episodic with at least two major events that
  occurred recently ($<$~28 years). The discrete mass-loss events
  together with the large turbulent width of the gas ($dv$=10--12 km
  s$^{-1}$) are consistent with the predictions of the Jet-Bow shock
  outflow and the wide-angle wind model. The CO gas/solid column
  density ratio of 20--100 in the line-of-sight confirms that the
  circumstellar environment of LLN~19 is warm. We also derive a
  $^{12}$C/$^{13}$C ratio of 67 $\pm$ 3, consistent with previous
  measurements in local molecular clouds but not with the higher
  ratios found in the envelope of other young stellar objects.
\end{abstract}


%
\section{Introduction}\label{iras08470_intro}

The gas dynamics of the inner surroundings of low- and high-mass young
stellar objects (YSOs) is a complex interplay between accretion,
settling and ejection (e.g.,
\citealt{Stahler2005fost.book.....S,Arce2007prpl.conf..245A}).
The gas and dust of the collapsing envelope are accreting onto the
YSOs or settling onto discs in Keplerian rotation.  Meanwhile,
powerful jets and winds remove the excess angular momentum dispersing
the envelope and allowing the system to contract. The release of
gravitational energy and the radiation of the YSOs heat the inner
circumstellar gas to a few hundred to a few thousand K.
Mid-infrared observations best probe these regions for two reasons.
First, dust grains absorb and scatter less in the infrared than in the
visible.  Second, atoms and molecules at a few hundred Kelvin emit
preferably in the mid-infrared.

Molecular ro-vibrational transitions trace the large range in density,
temperature and ultraviolet radiation field encountered in
circumstellar discs and envelopes.  While rotational lines in the
millimeter wavelength range trace the cool outer disc and envelope
($T<$~100~K), the mid-infrared fundamental emission ($v=1\rightarrow
0$) and absorption ($v=1\leftarrow 0$) probe the warm part
($T$=100--900 K) and the overtone emissions ($\Delta v$=2) in the
near-infrared the hot dust free region at $T>$1500 K
\citep{Bik2004A&A...427L..13B,Thi2005A&A...430L..61T}. CO
ro-vibrational lines seen in absorption against the continuum
generated by warm dust can probe cold as well as warm gases along the
line-of-sight.
 
Fundamental absorption lines of CO, the second most abundant molecule
after H$_2$, are detected in high-mass young stellar objects (e.g.,
\citealt{Mitchell1990ApJ...363..554M};
\citealt{Scoville1983ApJ...275..201S}). The line profiles indicate the
presence of quiescent cold gas located in the envelope as well as
outflowing warm gases. Ro-vibrational CO gas lines in emission and in
absorption have also been seen in the envelope (e.g,
\citealt{Boogert2002ApJ...568..761B};
\citealt{Pontoppidan2002A&A...393..585P,Pontoppidan2003A&A...408..981P,Brittain2007ApJ...659..685B})
or in the inner disc \citep{Lahuis2006ApJ...636L.145L} of low-mass
embedded young stellar objects. Besides CO, absorption lines of
formaldehyde \citep{Roueff2006A&A...447..963R}, C$_2$H$_2$, HCN, OCS,
NH$_3$ \citep{Evans1991ApJ...383..674E}, $^{13}$C$^{12}$CH$_2$,
CH$_3$, NH$_3$, HNCO, CS \citep{Knez2009ApJ...696..471K}, and methane
\citep{Boogert2004ApJ...615..344B} are observed towards high-mass YSOs
with high-dispersion spectrometers from the ground.

In this paper, we present an analysis of the CO fundamental absorption
lines toward LLN~19 (IRAS 08470--4321), an intermediate-mass YSO
located in the Vela Molecular cloud D at an estimated distance of
700~pc \citep{Liseau1992A&A...265..577L}. The luminosity of the object
($L_*$=1600~L$_{\odot}$) corresponds to a $\sim$~6-7~M$_{\odot}$
protostar \citep{Palla1993ApJ...418..414P}. LLN~19 is a deeply
embedded ($A_V\sim$~45~mag) YSO with moderate water and CO ice
abundance in its line-of-sight \citep{Thi2006}. Millimeter continuum
emission shows that LLN~19 is surrounded by 3.5 M$_{\odot}$ of gas and
dust in a 24~$\arcsec$ beam. Part of this matter likely forms a disc
and the rest remains in the envelope \citep{Massi1999A&AS..136..471M}.
A molecular outflow is detected in the $^{12}$CO $J=1 \rightarrow 0$,
C$^{17}$O $J=1 \rightarrow 0$ and CS $J=2 \rightarrow 1$ maps of
LLN~19 \citep{Wouterloot1999A&AS..140..177W}. The total mass of the
outflow reaches 55 M$_{\odot}$ if a kinematic distance of 2.24 kpc is
adopted
\citep{Wouterloot1999A&AS..140..177W}. \citet{Lorenzetti2002ApJ...564..839L}
found shocked H$_2$ emission around LLN~19 but could not definitively
show that LLN~19 is the source of the outflow. Observation
  evidence for the episodic nature of large scale molecular outflows
  is mounting in low- and high-mass YSOs (e.g.,
  \citealt{Arce2001ApJ...554..132A}). However, evidence for episodic
  warm inner winds exists primarily for high-mass YSOs from CO
  absorption studies (e.g.,
  \citealt{Mitchell1991ApJ...371..342M}). Our work extends the study
  of inner warm winds to lower luminosity objects.

We describe the observation and data reduction procedures in
Sect.~\ref{iras08470_obs} and give a first analysis of the results in
Sect.~\ref{iras08470_results}. We analyze the data using a wind model
for the {\em ISAAC} data and a wind-envelope model for the {\em
  CRIRES} data in Sect.~\ref{wind_model} and
Sect.~\ref{wind_envelope_model} and discuss the implications of our
analysis on the gas and ice content in the environment of LLN~19 in
Sect.~\ref{iras08470_discussion}. Finally, we conclude in
Sect.~\ref{iras08470_conclusion}.

\section{Observation and data reduction}\label{iras08470_obs}

We observed LLN~19 (IRAS 08470--4321) at a spectral resolving power of
$R\approx~10,000$ (resolution of 30 km s$^{-1}$) by setting the slit
width to 0.3\arcsec with the {\em Infrared Spectrometer And Array
  Camera} (ISAAC) mounted on the {\em Very Large Telescope}-UT1
(VLT-UT1) as one of the targets of a ESO large programme to study the
ice and gas around young stellar objects
\citep{vanDishoeck2003Msngr.113...49V}. The slit width corresponds to
a physical distance of 210~AU at the distance of LLN~19 (700~pc).
We adopted the distance of 700~pc derived by
  \cite{Liseau1992A&A...265..577L} over the kinematic distance of 2.24
  kpc. The closer distance is consistent with the YSO LLN~19 being a
  member of the Vela Molecular Cloud. We obtained a $M$-band spectrum
from 4.5 to 4.8 $\mu$m of LLN~19.  We also observed standard stars at
same airmass with the same instrument settings. We restrict our
analysis to the 4.55--4.8 $\mu$m region because the signal-to-noise
ratio is too low below 4.55 $\mu$m due to strong atmospheric CO$_2$
absorption.  The detector for 3-5~$\mu$m observations is a $\mathrm
1024\times1024$ Aladdin InSb array.  We reduced the data using an
in-house software written in the Interactive Data Language ({\em
  IDL}).  The data reduction procedure is standard for infrared
observations (bad pixel maps, flat-fielding, distortion correction,
etc...).  Because of the scarcity of halogen lamp lines in the
$M$-band, the spectrum was wavelength calibrated by comparing the
strong atmospheric absorption lines with high signal-to-noise ratio
spectra of the atmosphere above Paranal measured by the ESO staff.
The uncertainties in transmission are dominated by systematics from
the telluric line removal procedure
\citep{Pontoppidan2005hris.conf..168P}.  The precision of the
wavelength calibration is $\sim$ 5 km s$^{-1}$, i.e. 1/6 of the
resolution $dv$=30 km s$^{-1}$, over the entire spectrum.  Details on
the data reduction procedures are described in
\cite{Pontoppidan2003A&A...408..981P}. At the time of the observation
(7/1/2001), $v_{\mathrm{obs}}$=$v_{\mathrm{LSR}}$(LLN~19)+4.2 km
s$^{-1}$.

Subsequent high-resolution observations at $R$=100,000 ($dv$=3 km
s$^{-1}$) of LLN~19 were obtained with the Cryogenic Infrared Echelle
Spectrograph ({\em CRIRES}) at the VLT on April 26$^{\mathrm{th}}$
2007. {\em CRIRES} provides a resolving power of up to 10$^5$ in the
spectral range from 1 to 5$\mu$m when used with a 0.2 \arcsec
slit. The simultaneous spectral coverage between 4.64 and 4.85 $\mu$m
was achieved using a mosaic of four Aladdin III InSb arrays, which
combine into an effective 4096 x 512 focal plane detector array in the
focal plane. We used the Adaptive Optics system MACAO
(Multi-Applications Curvature Adaptive optics) to optimize the
signal-to-noise ratio and the spatial resolution. In our observation,
the signal-to-noise ratio reaches $\sim$~180. As for the {\em ISAAC}
observations, chopping and nodding along the slit were used and
standard stars were observed at the same airmass to correct for
telluric absorption lines. The {\em CRIRES} data were reduced using
standard procedures that includes flat-field correction, detector
non-linearity correction, and linearisation of the spectra in both the
dispersion- and cross-dispersion directions. We used a telluric model
to wavelength-calibrate the spectrum and to blank-out the telluric
absorptions after division by the standard. The width of the continuum
is 0.35 \arcsec, consistent with the AO-corrected spatial resolution.
The {\em CRIRES} data are part of a large programme that observed
$\sim$100 YSOs and protoplanetary discs
\citep{Pontoppidan2008ApJ...684.1323P}. A handful of the observed
  YSOs show high velocity CO absorption features that can be ascribed
  to winds. In particular, the spectrum of LLN~19 has the deepest CO
  absorption features.

\section{Observational results}\label{iras08470_results}

\begin{figure*}
     \centering
     \includegraphics[width=17cm]{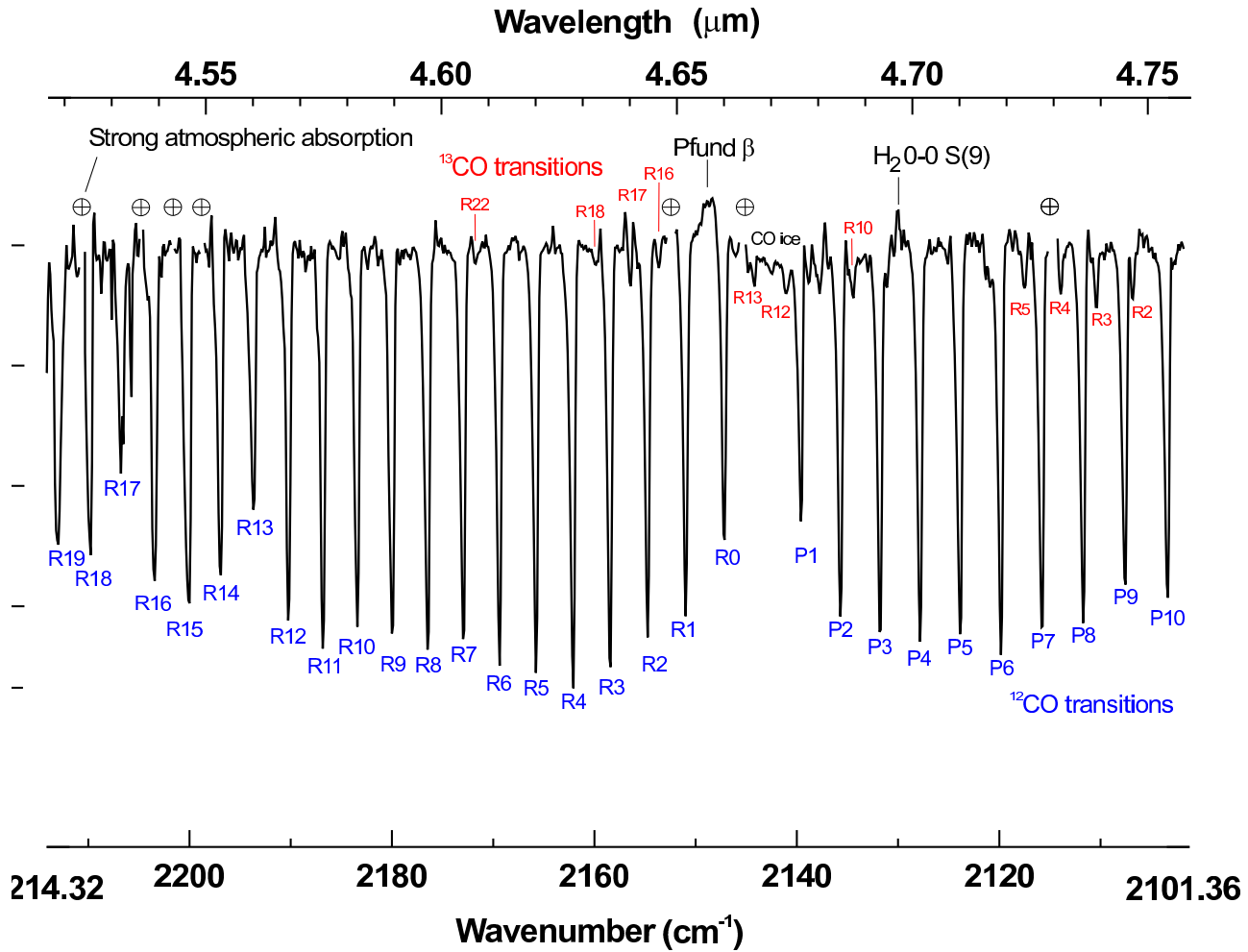}
   \caption{{\em VLT-ISAAC} $M$-band spectrum of LLN~19. The positions of the $^{12}$CO and $^{13}$CO lines are labelled. Wavelength regions with poor
     atmospheric transmission are omitted and labelled with a $\oplus$
     symbol.
\label{fig_co_labels}}
\end{figure*}

The continuum subtracted {\em ISAAC} spectrum in
Fig.~\ref{fig_co_labels} is dominated by deep gas-phase $^{12}$CO
ro-vibrational $v=1\leftarrow 0$ $R$-branch ($J+1\leftarrow $,
$J'=J+1$, $J"=J$) and $P$-branch ($J+1 \rightarrow J$) absorption
lines. A shallow absorption feature due to solid CO is also present in
the spectrum.

\begin{figure*}
     \centering
     \includegraphics[width=18cm]{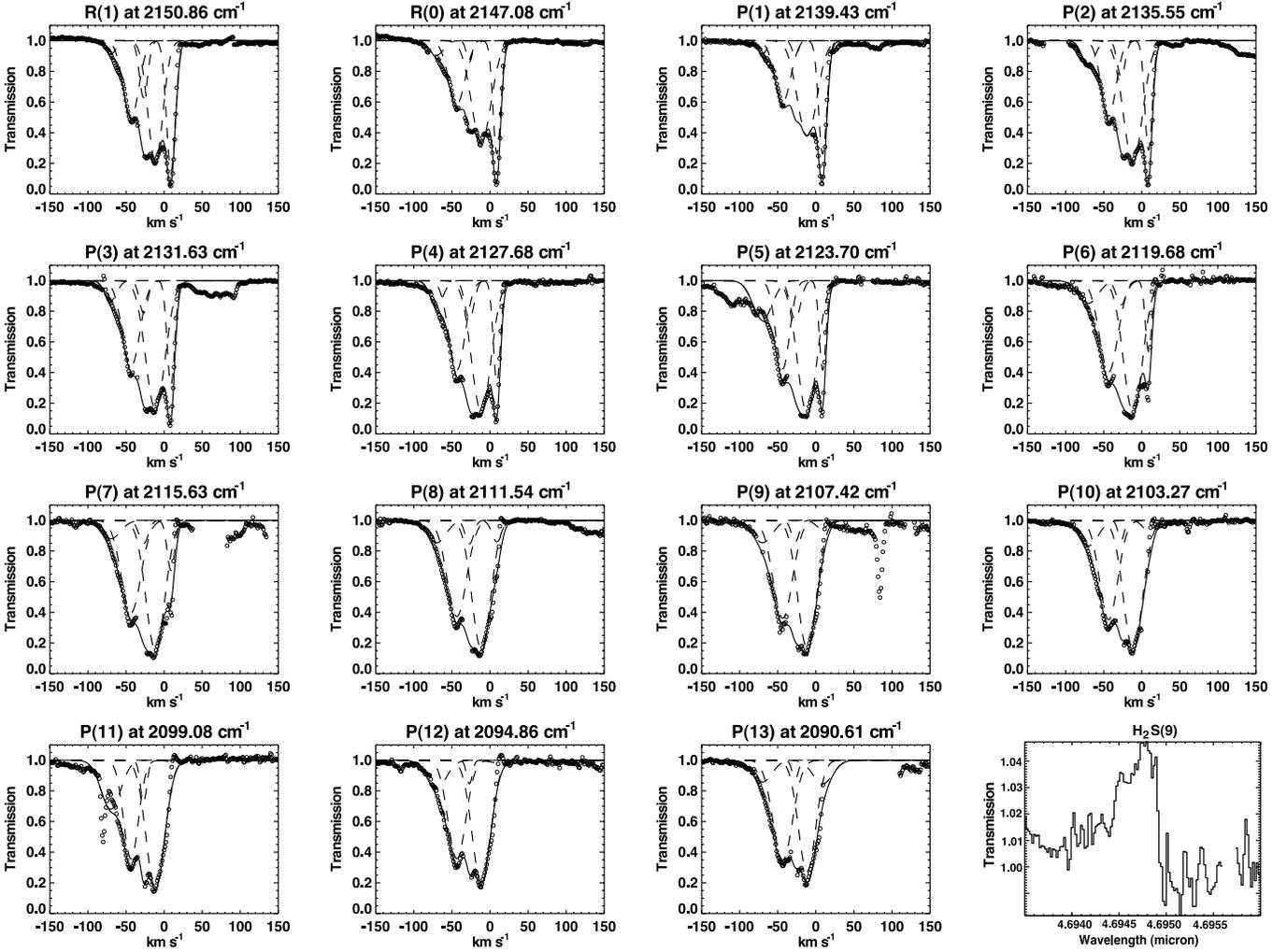}
   \caption{Absorption lines in open circles towards LLN~19 observed with {\em VLT-CRIRES} at a resolution of 3 km s$^{-1}$. Wavelength regions with poor atmospheric transmission are omitted. The long-dashed lines
are the substructure fits and the continuous lines are the total fits, which are the sum of all the Gaussian sub-components.The lower-right panel shows the tentatively detected H$_2$ $S(9)$ line at 4.6947 $\mu$m by {\em VLT-CRIRES}.
\label{fig_muliple_co_crires}}
\end{figure*}

$^{12}$CO $R$(0) and $R$(1), and $P$-branch transitions from 1 to 13 have
been observed with {\em CRIRES} and a sample of those absorption
lines is shown in Fig.~\ref{fig_muliple_co_crires}. The absorption lines
show multiple substructures, which correspond to the blue wing absorption
seen in the lower-resolution {\em ISAAC} spectrum (Fig.~\ref{fig_profile}).
Such data are available only for the lower $J$ lines, probing the cooler gas.

\subsection{CO ice feature}\label{iras08470_COice}

A weak feature centred around 4.675~$\mu$m (2140 cm$^{-1}$)
corresponding to CO-ice is detected. The total solid CO column density
is (4.3 $\pm$ 0.6) $\times$ 10$^{16}$ cm$^{-2}$ using the band strength of
$A$(CO)= 1.1 $\times$
10$^{-17}$ cm molec$^{-1}$ at 14~K
(\citealt{Jiang1975JChPh..62.1201J};
\citealt{Schmitt1989ApJ...340L..33S};
\citealt{Gerakines1995A&A...296..810G}) to convert the integrated absorption to column density. 
The detailed analysis of the CO ice components in relation to other YSOs in the Vela cloud is provided by \citet{Thi2006}.
\subsection{Gaseous components}

\subsubsection{ISAAC data}

We first focus on the {\em ISAAC} spectrum. The LLN~19 spectrum is
particularly rich in gas-phase lines.  We detect the hydrogen
recombination emission line $\mathrm{Pf}\beta$, which is an indicator
of accretion/wind activity and dense gases ($n_{\mathrm{e}}\sim$
10$^{8}$ cm$^{-3}$), at 4.653 $\mu$m. The broad wings ($\Delta
V$=300-500~km~s$^{-1}$) of the line can be ascribed to a stellar/disc
wind, especially in low- and intermediate-mass stars surrounded by
discs (e.g.\ \citealt{Carmona2005A&A...436..977C}). For high-mass
stars, the stellar wind is driven by radiative pressure exerted on the
ionic lines. We tentatively detect the pure-rotational H$_2$
  $0-0$ $S(9)$ line at 4.695~$\mu$m with an equivalent width of
  (27~$\pm~$12)~$\times$~10$^{-3}$ cm$^{-1}$ (1 $\sigma$). The line is
  also seen in the {\em CRIRES} data (see the lower-right panel of
  Fig.~\ref{fig_muliple_co_crires}) with an equivalent width of
  (11.4~$\pm~$3.5)~$\times$~10$^{-3}$ cm$^{-1}$. The detection of the
H$_2$ line in emission, whose energy level lies at 5005.73~K,
testifies to the presence of hot molecular gas.

\begin{figure}
  \centering
  \resizebox{\hsize}{!}{\includegraphics[]{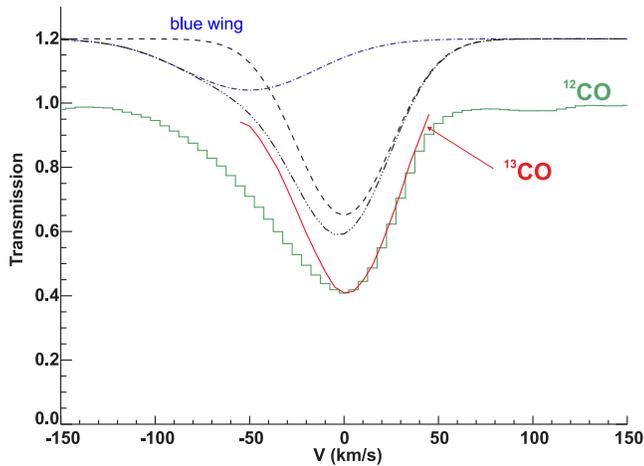}}
  \caption{Comparison of the $^{12}$CO and $^{13}$CO average {\em
      ISAAC} profiles. The profiles are averages of all the detected
    lines to increase the signal-to-noise ratio. The profiles are
    corrected for the Earth velocity at the day of the observation and
    are normalised to an absorption depth of 0.4 to ease the
    comparison. The model decompositions are shifted by +0.2 for
    clarity. Both mean profiles were shifted by -9 km
    s$^{-1}$.\label{fig_profile}}
\end{figure}

$^{12}$CO $R-$ and $P$-branch lines, up to $J$=10 and $J$=19
respectively, are clearly seen (Fig.~\ref{fig_co_labels}).  The lines
are up to 60~\% deep at the spectral resolution of 30 km
s$^{-1}$. Columns 1 and 2 of Table~\ref{CO12LineTable} contain the
rest laboratory and observed wavenumbers. The quiescent cold gas
around LLN~19 is located at $v_{\mathrm{LSR}}$(mm)~=~12~km~s$^{-1}$
from millimeter observations of $^{12}$CO $J=1 \rightarrow 0$ in a
46$\arcsec$ beam \citep{Wouterloot1999A&AS..140..177W}. By contrast,
the CO gas seen in the mid-infrared is moving by an average
$v_{\mathrm{LSR}}$~=~-25~km~s$^{-1}$ ($\pm$~5~km~s$^{-1}$). The line
velocities are not randomly scattered around the average velocity
shift value. The low-$J$ lines have line widths smaller than the
high-$J$ lines, suggesting that quiescent cold gas may also be present
in the line-of-sight, but the spectral resolution is too low to
resolve it from the outflowing gas. The line profiles are not
symmetric and can be decomposed into two Gaussian components: a deep
component centred at $v_{\mathrm{LSR}}$~=~-25~$\mp$~5~km~s$^{-1}$ and
Full Width at Half Maximum $FWHM$= 33 km s$^{-1}$ and a broad
shallower component at $v_{\mathrm{LSR}}$~=~-75~$\mp$~5~km~s$^{-1}$
with $FWHM$~=~59 km s$^{-1}$. Fig.\ \ref{fig_profile} displays a
normalised absorption line profile averaged over all the lines and its
decomposition into the two components. The lines do not broaden with
increasing quantum rotational number for the transition. The
decomposition is arbitrary but allows a simple analysis of the two
components.

In the rest of the paper, we will refer to the main (at -25 km
s$^{-1}$) and blue (at -75~km~s$^{-1}$) component. At the resolution
of $\sim$~30~km~s$^{-1}$, the main and blue CO absorption lines are
marginally resolved; the depth of the absorption lines indicates that
the intrinsic width $\Delta v$ is $\sim
\sqrt{FWHM^2-Resolution^2}=\sqrt{33^2-30^2}~=~13.7$~km~s$^{-1}$ if the
main absorption profile is caused by a single component. High velocity
blueshifted gas is commonly found towards high-mass
\citep{Mitchell1991ApJ...371..342M} and low-mass YSOs (e.g.,
\citealt{Lahuis2006ApJ...636L.145L}).

Weaker $^{13}$CO $R$-branch absorption lines are detected up to $J$=22
as well and are labelled in Fig.~\ref{fig_co_labels}. The blue part of
the average $^{13}$CO profile is contaminated by other features,
mostly by the wings of the $^{12}$ CO feature. The
$^{13}$CO lines are seen at the same velocity than their $^{12}$CO
counterpart. However, the weakness of the $^{13}$CO lines forbids an
accurate determination of the shifts.

The equivalent widths of the main and blue component are given in
Tables \ref{CO12LineTable} and \ref{CO13LineTable}. The normalised
equivalent width ratios
$W_{\tilde{\nu}}/{\tilde{\nu}}=W_\lambda/\lambda$ between $^{12}$CO
and $^{13}$CO lines increase with increasing rotational quantum number
but remain much lower than the interstellar isotopic value of
$^{12}$C/$^{13}$C~=~60--65, suggesting that the $^{12}$CO lines are
optically thick.

%
\begin{figure*} 
  \centering
\includegraphics[height=9.5cm,width=18cm]{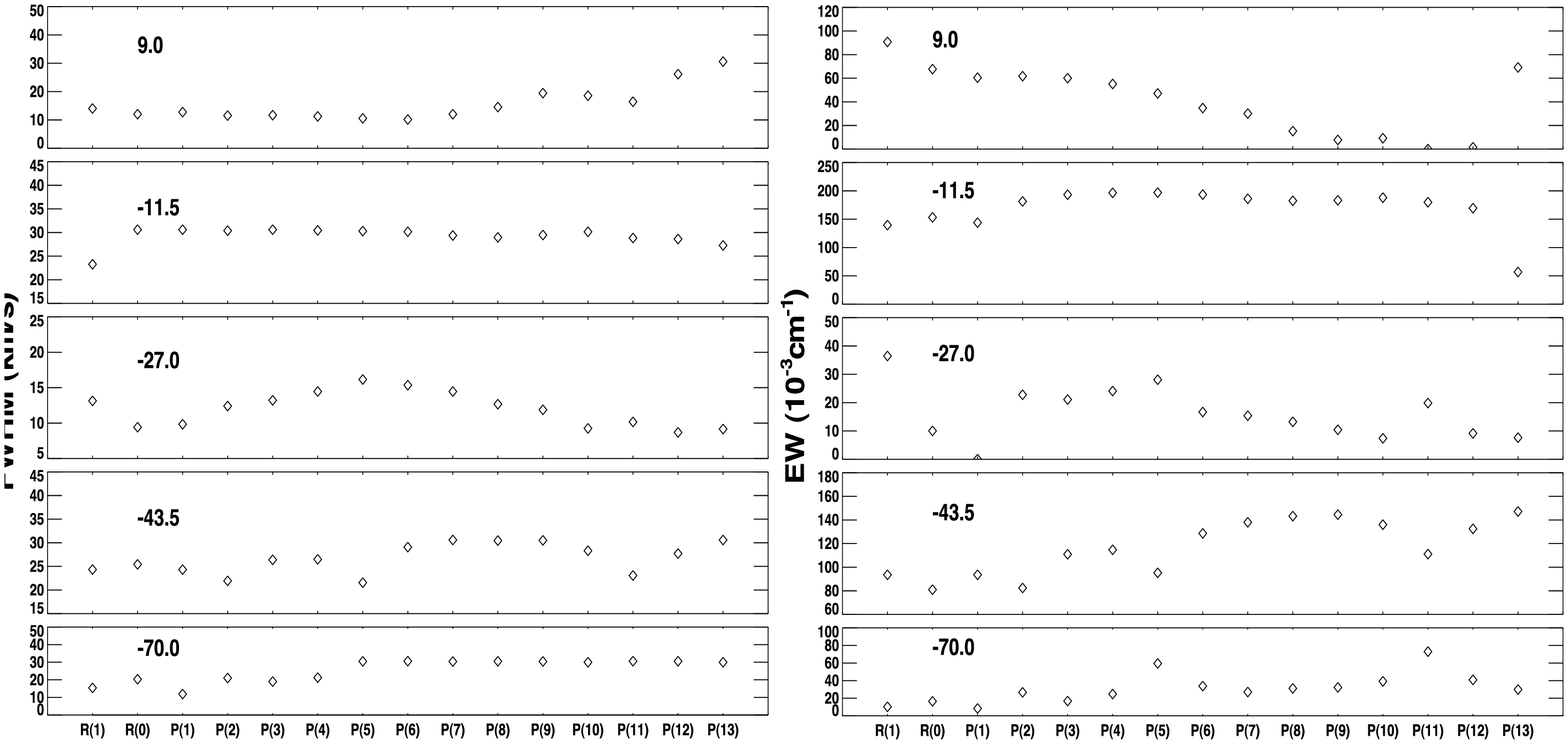}
   \caption{Gaussian width (left panel) and equivalent width (right panel) for each individual substructure and each transitions (x axis) for the {\em CRIRES}. The number in each panel indicates the velocity of the substructures. The main causes of uncertainties are missing data due to strong telluric absorptions. \label{fig_fwhm_EW_crires}}
\end{figure*}
%

\subsubsection{CRIRES data}
  
The {\em CRIRES} spectra show that the two components seen in the {\em
  ISAAC} spectrum actually consist of the absorption by four clumps at
9,~-11.5,~-27,~-43.5, and -70~km~s$^{-1}$. The components at 9,~-11.5,
and~-27 km~s$^{-1}$ correspond to the main component in the {\em
  ISAAC} data while the -43.5 and -70~km~s$^{-1}$ components correspond to
the blue wing. The absorption at 9~km~s$^{-1}$ may be saturated at the
resolution of 3~km~s$^{-1}$. As a result the measured widths of the
9~km~s$^{-1}$ feature may be broader than the intrinsic widths.

We modelled the absorption lines by a combination of four Gaussians
with fixed velocities $v_i$ (9,~-11.5,~-27,~-43.5, and
-70~km~s$^{-1}$). We performed our Gaussian decomposition in the
  transmission domain since we are interested in the equivalent widths
  of each component. The fitting function has the form:
\begin{equation}
f(v)=1-\sum_i A_i \exp\left(-\left(\frac{v-v_i}{\delta v_i}\right)^2\right)
\end{equation}
where $A_i$ and $\delta v_i$, the widths and amplitudes of the
Gaussians, are found by the fitting routine.  The width and
amplitudes of the Gaussians are free parameters of the fits.  We used
the Shuffle Complex Evolution optimization code (see
Sect.~\ref{Model_description}) to determine the best fits.  The fit
for each individual line is shown in Fig.~\ref{fig_muliple_co_crires}
whereas the FWHM of the Gaussians and the equivalent widths are shown
in Fig.~\ref{fig_fwhm_EW_crires}.

The largest equivalent widths ($EW>$~0.1~cm$^{-1}$) correspond to the
outflowing component at -11 km s$^{-1}$.  The quiescent envelope
component at 9 km s$^{-1}$ shows decreasing equivalent widths with
increasing $J$ values, testifying of a cold component. The large EW
for $P$(13) in the {\em CRIRES} data for the component at 9 km
  s$^{-1}$ is most likely an artifact caused by missing data. The gas
equivalent widths do not show similar clear trend with increasing $J$
values.

The average FWHM of the Gaussians is consistent with an intrinsic
turbulent width of $\sim$~13.7~km~s$^{-1}$ inferred from the {\em
  ISAAC} data. The parameters will be analysed in the following
sections.

\subsubsection{Comparison between ISAAC and CRIRES data}

\begin{figure}
  \centering
  \resizebox{\hsize}{!}{\includegraphics[]{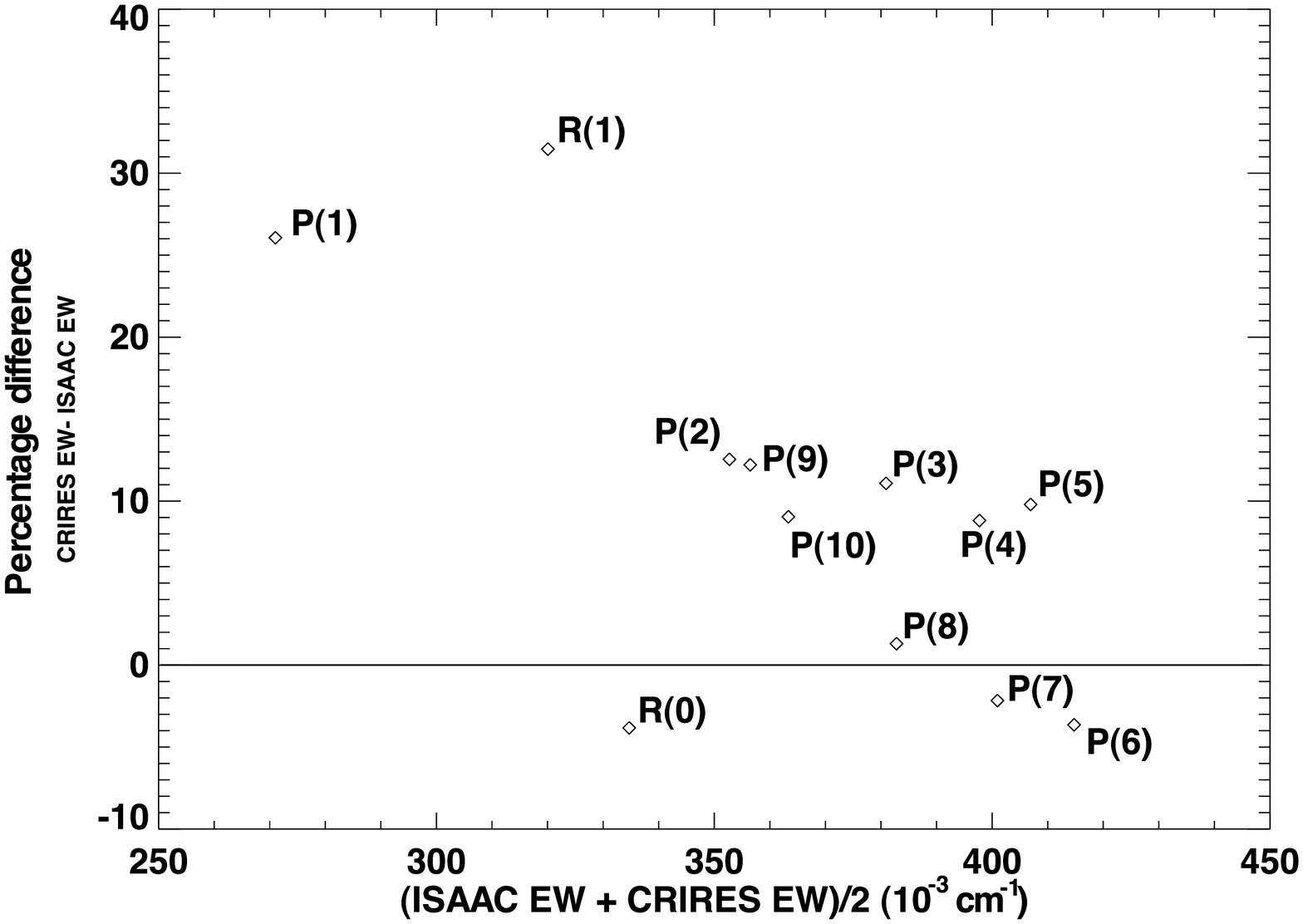}}
  \caption{Plot of percentage difference versus mean value between
    equivalent widths derived from the {\em ISAAC} and the {\em
      CRIRES} observations. The systematic differences are below 20\%
    except for the $P(1)$ and $R(1)$ transitions
    ($\sim$~30\%).\label{fig_ew_comparison}}
\end{figure}

The measured equivalent widths derived from {\em ISAAC} and {\em
  CRIRES} data are compared in
Fig.~\ref{fig_ew_comparison}. Discrepancies are small for most
transitions ($<$15\%) except for $P$(1) and $R$(0). Generally the {\em
  CRIRES} equivalent widths are larger than the {\em ISAAC} equivalent
widths. The lower resolution {\em ISAAC} data may suffer from
systematic errors in the determination of the continuum, especially
when the $^{12}$CO blue-wing absorptions reach the wavelengths of the
$^{13}$CO absorption. The large differences for $P$(1) and $R$(0) may stem
from the incomplete profiles in the {\em CRIRES} data-set at those
transitions due to strong telluric absorptions. In case of strong
telluric absorptions, the data are interpolated to compensate for the
missing data for the determination of the equivalent width and this
procedure can introduce significant systematic errors, in our case an
overestimate of the actual absorption (see
Fig.~\ref{fig_muliple_co_crires}). Strong telluric absorption features
also introduce errors in the {\em ISAAC} equivalent widths since the
raw {\em ISAAC} data are ratioed by a standard star
\citep{Pontoppidan2005hris.conf..168P}.  Alternatively the choices
made in the fitting of the CO ice absorption feature affect the
continuum flux around the $P$(1) and $R$(0) lines more strongly in the
{\em ISAAC} spectrum than in the {\em CRIRES} spectrum (see
Fig.~\ref{fig_co_labels}). Therefore the $P$(1) and $R$(0) equivalent
widths measured by {\em ISAAC} may be underestimated. In summary, the
differences in EW between the two datasets can probably be ascribed to
systematic errors in interpolating missing data and in cancelling
telluric absorptions.

%

\begin{table}
\centering
\begin{flushleft}
\caption{$^{12}$CO $v=1 \leftarrow 0$ $R$- and $P$-branch line positions and equivalent widths of the middle component (mc) and blue component (bc) toward LLN~19.\label{CO12LineTable}}
\begin{tabular}{lllll}
\hline 
\hline
\noalign{\smallskip} 
\multicolumn{1}{c}{Line} &  \multicolumn{1}{c}{${\tilde{\nu}}_{lab}$} & \multicolumn{1}{c}{${\tilde{\nu}}_{obs}$(mc)} & \multicolumn{1}{c}{$W_{\tilde{\nu}}$(mc)}& \multicolumn{1}{c}{$W_{\tilde{\nu}}$(bc)} \\
   &  \multicolumn{2}{c}{(cm$^{-1}$)} & \multicolumn{2}{c}{(10$^{-3}$ cm$^{-1}$)}\\
\noalign{\smallskip} 
\hline
\noalign{\smallskip}
$ R(0)$ &2147.081 & 2147.19 & 208.3 $\pm$  1.6 & \ \ 61.4 $\pm$  0.3 \\
$ R(1)$ &2150.856 & 2150.96 & 263.4 $\pm$ 14.3 & \ \ 77.7 $\pm$  3.0 \\
$ R(2)$ &2154.596 & 2154.72 & 279.7 $\pm$ 10.9 & \ \ 82.3 $\pm$  2.3 \\
$ R(3)$ &2158.300 & 2158.43 & 300.5 $\pm$  2.5 & \ \ 88.7 $\pm$  0.5 \\
$ R(4)$ &2161.969 & 2162.07 & 306.9 $\pm$  0.5 & 139.9 $\pm$  0.2 \\
$ R(5)$ &2165.601 & 2165.74 & 296.8 $\pm$  0.7 & 135.2 $\pm$  0.2 \\
$ R(6)$ &2169.198 & 2169.34 & 292.5 $\pm$  6.9 & 133.3 $\pm$  2.3 \\
$ R(7)$ &2172.759 & 2172.87 & 273.5 $\pm$  5.2 & 145.9 $\pm$  2.0 \\
$ R(8)$ &2176.284 & 2176.42 & 281.3 $\pm$  7.3 & 171.4 $\pm$  3.2 \\
$ R(9)$ &2179.772 & 2179.93 & 279.4 $\pm$  8.9 & 123.6 $\pm$  2.8 \\
$R(10)$ &2183.274 & 2183.38 & 275.6 $\pm$  4.8 & \ \ 81.1 $\pm$  1.0 \\
$R(11)$ &2186.639 & 2186.84 & 291.9 $\pm$  6.3 & \ \ 85.9 $\pm$  1.3 \\
$R(12)$ &2190.010 & 2190.24 & 271.9 $\pm$  1.2 & \ \ 80.0 $\pm$  0.3 \\
$R(13)$ &2193.359 & 2193.66 & 192.0 $\pm$  1.2 & \ \ 70.6 $\pm$  0.3 \\
$R(14)$ &2196.664 & 2196.88 & 239.8 $\pm$  1.0 & \ \ 70.6 $\pm$  0.2 \\
$R(15)$ &2199.931 & 2200.03 & 232.4 $\pm$  3.9 & 191.6 $\pm$  2.3 \\
$R(16)$ &2203.161 & 2203.42 & 231.6 $\pm$  5.3 & 107.9 $\pm$  1.8 \\
$R(17)$ &2206.354 & 2206.53 & 137.5 $\pm$ 27.4 & 149.5 $\pm$ 21.3 \\        
$R(18)$ &2209.509 & 2209.80 & 213.9 $\pm$ 21.4 & 133.2 $\pm$  9.5 \\
$R(19)$ &2212.626 & 2212.94 & 218.9 $\pm$ 31.9 & \ \ 96.6 $\pm$ 10.1 \\
\noalign{\smallskip} 
\hline
\noalign{\smallskip} 
$ P(1)$ &2139.426 & 2139.59 & 182.0 $\pm$  9.6 & \ \ 53.7 $\pm$  2.0 \\
$ P(2)$ &2135.546 & 2135.77 & 255.3 $\pm$  1.9 & \ \ 75.3 $\pm$  0.4 \\
$ P(3)$ &2131.632 & 2131.74 & 272.3 $\pm$  1.3 & \ \ 87.5 $\pm$  0.3 \\
$ P(4)$ &2127.684 & 2127.84 & 277.8 $\pm$  3.5 & 102.4 $\pm$  0.9 \\
$ P(5)$ &2123.699 & 2123.86 & 278.4 $\pm$  1.8 & 108.6 $\pm$  0.5 \\
$ P(6)$ &2119.681 & 2119.86 & 286.9 $\pm$  2.5 & 135.4 $\pm$  0.8 \\
$ P(7)$ &2115.629 & 2115.78 & 267.3 $\pm$  2.1 & 138.0 $\pm$  0.8 \\
$ P(8)$ &2111.543 & 2111.71 & 255.7 $\pm$ 12.0 & 124.6 $\pm$  4.2 \\
$ P(9)$ &2107.424 & 2107.57 & 229.7 $\pm$  6.7 & 105.0 $\pm$  2.2 \\
$P(10)$ &2103.270 & 2103.40 & 238.1 $\pm$  3.7 & 108.8 $\pm$  1.2 \\
\noalign{\smallskip} 
\hline
\noalign{\smallskip} 
\end{tabular}

\end{flushleft}
\begin{list}{}{}
\item[$^\mathrm{a}$] 3~$\sigma$ fitting error.
\end{list}
\end{table}

\begin{table}
\centering
\begin{flushleft}
\caption{$^{13}$CO $v=1\leftarrow 0$ $R$-branch line positions and equivalent widths toward LLN~19\label{CO13LineTable}}
\begin{tabular}{lllll}
\hline 
\hline
\noalign{\smallskip} 
\multicolumn{1}{c}{Line} &  \multicolumn{1}{c}{${\tilde{\nu}}_{lab}$} & \multicolumn{1}{c}{${\tilde{\nu}}_{obs}$(mc)} & \multicolumn{1}{c}{$W_{\tilde{\nu}}$(mc)}& \multicolumn{1}{c}{$W_{\tilde{\nu}}$(bc)} \\
   &  \multicolumn{2}{c}{(cm$^{-1}$)} & \multicolumn{2}{c}{(10$^{-3}$ cm$^{-1}$)}\\
\noalign{\smallskip} 
\hline
\noalign{\smallskip} 
$R(2)$  & 2106.903 & 2106.82 & 42.5 $\pm$ 4.5$^\mathrm{a}$ & $<$5.9\\
$R(3)$  & 2110.447 & 2110.42 & 48.5 $\pm$ 4.5 & ... \\
$R(4)$  & 2113.958 & 2113.85 & 39.0 $\pm$ 4.5 & $\sim$9.6\\
$R(5)$  & 2117.436 & 2117.57 & 34.4 $\pm$ 4.5 & ... \\
$R(10)$ & 2134.318 & 2134.38 & 27.4 $\pm$ 4.5 & $\sim$21.7\\
$R(12)$ & 2140.833 & 2141.05 & 21.3 $\pm$ 4.5 & ...\\
$R(13)$ & 2144.039 & 2144.15 & 19.9 $\pm$ 4.5 & ...\\
$R(16)$ & 2153.447 & 2153.59 & 19.7 $\pm$ 4.5 & ... \\
$R(17)$ & 2156.514 & 2156.45 & 22.4 $\pm$ 4.5 & ... \\
$R(18)$ & 2159.545 & 2159.83 & 17.7 $\pm$ 4.5 & ... \\
$R(22)$ & 2171.317 & 2171.75 & 16.4 $\pm$ 4.5 & ... \\
\noalign{\smallskip} 
\hline
\end{tabular}
\end{flushleft}
\begin{list}{}{}
\item[$^\mathrm{a}$] formal 3 $\sigma$ error dominated by the flux
  calibration accuracy.
\end{list}
\end{table}
\section{Equivalent width analysis of the main component using {\em ISAAC} data}\label{cog_slab}

\subsection{Curve-of-growth analysis}~\label{cog_analysis}

We performed a classical curve of growth analysis of the {\em ISAAC}
data, which relates the measured equivalent width $W_{\tilde{\nu}}$
(in cm$^{-1}$) at wavenumber ${\tilde{\nu}}$ (in cm$^{-1}$) to the CO
column density $N_{J''}$ (in cm$^{-2}$) in the initial lower
rotational state $J''$ assuming a Gaussian profile with an
intrinsic line width called the Doppler parameter $b_D=
  \Delta v_{\mathrm{FWHM}}/ 2\log(2)^{1/2}$, where $\Delta
  v_{\mathrm{FWHM}}$ is the Full With at Half Maximum in km s$^{-1}$
  (e.g., \citealt{Spitzer1978ppim.book.....S}, Eq. 3.22), and a slab
  geometry.  We will focus on the {\em ISAAC} main component. The
equivalent widths $W_{\tilde{\nu}}$ and $N_{J''}$ are linearly
proportional only in the optically thin regime:

\begin{equation}
\frac{W_{\tilde{\nu}}}{{\tilde{\nu}}}=\left(1-\frac{S_{J''}}{I^B}\right)\frac{\pi e^2}{m_e c^2}\frac{f_{J'\leftarrow J''}N_{J''}}{{\tilde{\nu}}}
\label{equ_ew_slab}
\end{equation}

where $f_{J'\leftarrow J''}$ is the oscillator strength
(dimensionless) of the transition
\citep{Goorvitch1994ApJS...91..483G}, $I^B_\nu$ is the background
intensity and $S_{J''}$ is the transition source function in the
  extended region covered by the beam and the other symbols have
their usual meaning. We assume here that $S_{J''}/I^B_\nu\ll 1
$. Therefore the derived column densities are lower limits to the
actual values.

   \begin{figure}
     \centering
     \resizebox{\hsize}{!}{\includegraphics[]{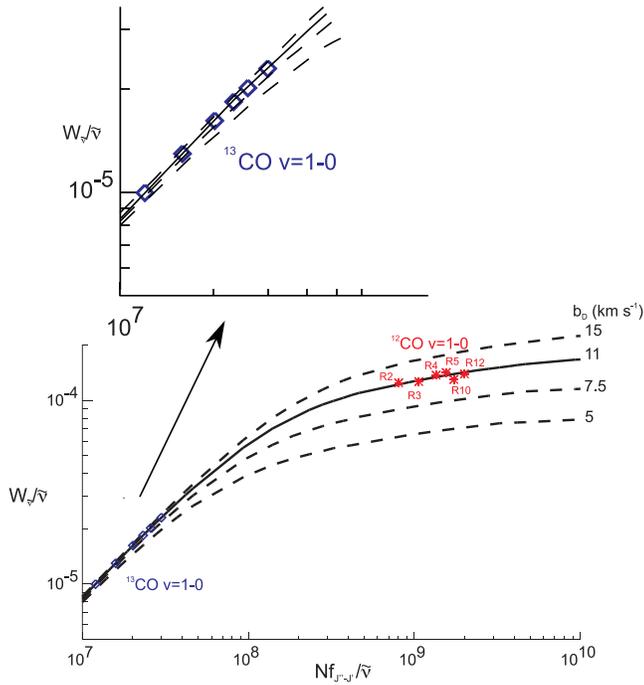}}
     \caption{Curve of growth. Only the $^{12}$CO equivalent widths
       (EW) are affected by saturation and thus can be used to
       constrain the Doppler parameter $b_{\mathrm{D}}$. The $^{13}$CO
       EW are on the linear part of the curve. The fit for the
       $^{13}$CO EW are also shown in the inset for better
       clarity. The best fit curve to the data is obtained with
       $b_{\mathrm{D}}$=11 $\pm$ 1.5 km s$^{-1}$ and $^{12}$C/$^{13}$C
       = 67 $\pm$ 3. \label{fig_cog}}
\end{figure}

Transitions that were detected in both $^{12}$CO and $^{13}$CO with
good signal-to-noise ratio in the {\em ISAAC} data are shown in
Fig.~\ref{fig_cog}. The vertical axis corresponds to the observed
$W_{\tilde{\nu}}/{\tilde{\nu}}$, while the horizontal axis corresponds
to the different values of $N_{J''}$($^{13}$CO)$f_{J'\leftarrow
  J''}$($^{13}$CO) /${\tilde{\nu}}$ and
$N_{J''}$($^{12}$CO)$f_{J'\leftarrow
  J''}$($^{12}$CO)/${\tilde{\nu}}$. Theoretical curve-of-growths
computed assuming various values for the Doppler parameter $b_{D}$ are
overplotted in dashed lines. The column density in a given level $J''$
is computed by the equation
\begin{equation}
N_{J''}(^{13}{\mathrm{CO}}) = b_{D} \tau (^{13}{\mathrm{CO}})\tilde{\nu}/(1.497 \times 10^{-2} f_{J'\leftarrow
  J''}).
\end{equation}
Once the $\tau$($^{13}$CO) values are derived from the relationship
between $\tau$ and $W_{\tilde{\nu}}$ for each value of
$b_{\mathrm{D}}$ via the theoretical function $F(\tau)=0.5 c
(W_{\tilde{\nu}}/{\tilde{\nu}})/b_{\mathrm{D}}$
\citep{Spitzer1978ppim.book.....S}, the column density of $^{12}$CO at
level $J''$ is
\begin{equation}
N_{J''}(^{12}{\mathrm{CO}})= \zeta(^{12}{\mathrm C}/^{13}{\mathrm C}) N_{\mathrm{J''}}(^{13}{\mathrm{CO}}). 
\end{equation}
The ratio between the isotopologues $^{12}$CO/$^{13}$CO,
$\zeta(^{12}$C/$^{13}$C), is an adjustable parameter.  In
Fig.~\ref{fig_cog}, the $^{13}$CO data lie on the linear part of the
curves; therefore, the lines are optically thin whereas the $^{12}$CO
lines lie on the saturated part of the curve and are therefore
optically thick.

The best fit parameters were found by minimizing the difference
between the data and the empirical curve (least squares). First the
$\zeta(^{12}$C/$^{13}$C) parameter was varied. Varying $\zeta$ will only move
the $^{12}$CO points left or right in the curve-of-growth diagram. For
each $\zeta$ value $\chi^2$ values between the observed EWs and the
synthetic curves were determined by varying $b_{\mathrm{D}}$. The
determination of $\zeta(^{12}$C/$^{13}$C) and $b_{\mathrm{D}}$ is
independent on the actual CO column density and excitation.  The best
fit were obtained for $\zeta(^{12}$C/$^{13}$C)=67 $\pm$ 3 (1 $\sigma$) and
$b_{\mathrm{D}}$=11 $\pm$ 1.5 km s$^{-1}$ (1 $\sigma$). Since the EW
derived from {\em ISAAC} are strictly lower limits, we should consider
that $\zeta=67$ is also a lower limit. The errors were determined from the
$\chi^2$ surface.  The $\zeta(^{12}$C/$^{13}$C) value derived here is an
average between the $\zeta$ value in the warm wind and that in the cold
foreground cloud. The {\em CRIRES} spectra show that the main
component profile is a combination of multiple Gaussians. The $\zeta$
value of each individual velocity components can be derived using {\em
  CRIRES} data in future studies.  The average value of
$\zeta(^{12}$C/$^{13}$C) of 67 $\pm$ 3 is close to the local interstellar
value of 65--77 (e.g.,
\citealt{Bensch2001ApJ...562L.185B,Sheffer2007ApJ...667.1002S,Wilson1994ARA&A..32..191W}).

\subsection{Rotational diagram}
  
\subsubsection{Main component}  

\begin{figure}  
     \centering
     \resizebox{\hsize}{!}{\includegraphics[]{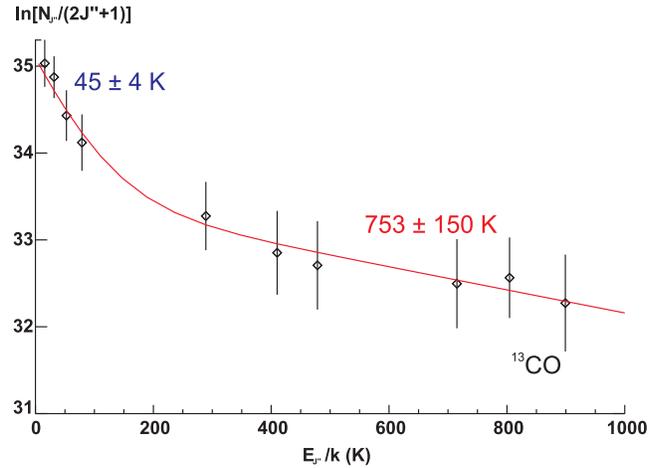}}
     \caption{Rotational diagram of $^{13}$CO for the main
       component. The fit by the 2-slab model is
       overplotted.\label{fig_rot_diag}}
\end{figure}

We used the optically thin $^{13}$CO data to construct the rotational
diagram based on the {\em ISAAC} data that is shown in
Fig.~\ref{fig_rot_diag} for the main component. In a rotational
  diagram for $R$-branch absorption lines the natural logarithm of the
  lower level column density of a transition $J' \leftarrow J''$
  divided by the degeneracy $(2J''+1)$, ($\log[N_{J''}/(2J''+1)]$) is
  plotted against the lower level energy $E_{J''}/k$ (in K).  If the
  level population is thermalised (LTE), the column density at a given
  level $J''$ is expressed as:
\begin{equation}
N_{J''} = N(\mathrm{CO})
(2J''+1)\frac{\exp(-E_{J''}/T_{kin})}{Q(T_{kin})}, 
\end{equation}
where $N(\mathrm{CO})$ is the column density of CO, $E_{J''}$ is the
energy of the $J''$th rotational level expressed in K,
$Q(T_{\mathrm{kin}}) \simeq kT_{kin}/B_0+1/3$ is a good approximation
of the partition function at the gas temperature $T_{kin}$ (equals to
the gas excitation temperature at high temperatures and densities) and
$B_0$ is the rotational constant with $B_0/k$= 2.765 and 2.644~K for
$^{12}$CO and $^{13}$CO respectively. The data points in a rotational
diagram can be fitted by a straight line for an isothermal medium at
LTE. The gas temperature is then given by the negative of the inverse
of the slope of the line. However, fits with a single temperature gas
to the observed values provide bad results (Fig.~\ref{fig_rot_diag}).
Possible non-LTE population was checked using the simple analytical
results of \cite{McKee1982ApJ...259..647M}.  High-$J$ levels in a gas
at non-LTE are underpopulated compared to a gas at LTE, which is not
seen in our data. Instead, the data are best fitted with the
combination of a cold and hot gas at LTE, similar to most young
stellar objects where CO fundamental absorption lines have been
observed.
\begin{table*}
\centering
\caption{Column densities and temperatures derived from the rotational diagram analysis. The upper part of
  the table corresponds to the parameters derived using the {\em ISAAC} data while in the lower part of the table corresponds to the parameters driven using the {\em CRIRES} data. The formal errors are 1$\sigma$ from the least squares fits. The $^{12}$CO column densities of the cold and warm components are derived using the optically thin $^{13}$CO data while the column densities in the blue wing were derived using the $^{12}$CO data. The fits to the wing data are poor, which explains the large errors in the derived temperatures and column densities. The column densities derived from the {\em CRIRES} equivalent widths are corrected for optical depth effects.
  \label{RotDiagTable}}
\begin{tabular}{llll}
  \hline 
  \hline
  \noalign{\smallskip} 
  \multicolumn{1}{c}{Component} & \multicolumn{1}{c}{$T$} & \multicolumn{1}{c}{$N$($^{12}$CO)}& \multicolumn{1}{c}{$N$($^{13}$CO)} \\ 
  & \multicolumn{1}{c}{(K)} & \multicolumn{1}{c}{(cm$^{-2}$)} & \multicolumn{1}{c}{(cm$^{-2}$)} \\
  \noalign{\smallskip} 
  \hline
  \noalign{\smallskip} 
  \multicolumn{4}{c}{{\em ISAAC data}}\\
  \noalign{\smallskip} 
  Cold main & 45~$\pm$~4  & (2.2~$\pm$~0.6)~$\times$~10$^{18\ a}$ & (3.3~$\pm$~0.1)~$\times$~10$^{16}$\\
  Warm main & 753~$\pm$~150 & (6.7~$\pm$~0.6)~$\times$~10$^{18\ a}$& (1.0~$\pm$~0.1)~$\times$ 10$^{17}$\\
  Cold blue wing      & 116$^{+53}_{-28}$ & 2.0$^{+1.7}_{-0.9}$~$\times$~10$^{17}$  & ...\\
  Warm blue wing      & 1338$^{+2298}_{-518}$ & 3.3$^{+5.4}_{-1.2}$~$\times$~10$^{17}$  & ...\\
  \noalign{\smallskip} 
  \hline
  \noalign{\smallskip} 
  \multicolumn{4}{c}{{\em CRIRES data}}\\
  \noalign{\smallskip} 
  at +9 km s$^{-1}$    & 54$^{+9}_{-7}$       &  1.4$^{+1.3}_{-0.8}$~$\times$~10$^{17}$ & ...\\
  at -11.5 km s$^{-1}$ & 214$^{+33}_{-24}$    &  5.2$^{+1.8}_{-1.2}$~$\times$~10$^{17}$ & ...\\
  at -27 km s$^{-1}$   & 397$^{+1050}_{-192}$ &  6.9$^{+183}_{-4.3}$~$\times$~10$^{16}$ & ...\\
  at -43.5 km s$^{-1}$ & 291$^{+66}_{-46}$    &  4.7$^{+2.1}_{-1.3}$~$\times$~10$^{17}$ & ...\\
  at -70 km s$^{-1}$   & 387$^{+128}_{-77}$    &  1.9$^{+0.7}_{-1.1}$~$\times$~10$^{17}$ & ...\\
  \noalign{\smallskip} 
  \hline
\end{tabular}
\begin{flushleft}
  $^a$ from $^{13}$CO using $^{12}$CO/$^{13}$CO=67.\ \\
\end{flushleft}
\end{table*}
The low-$J$
absorption lines are formed in a cold region ($T_{\mathrm{gas}}$=
45~K, $N$($^{13}$CO)$\simeq$ 3.3 $\times$ 10$^{16}$ cm$^{-2}$,
$N$($^{12}$CO)= 67 $\times$ $N$($^{13}$CO) $\simeq$ 2.2 $\times$
10$^{18}$ cm$^{-2}$ ) while the high-$J$ lines ($J\geq$10) are formed
in warm gas ($T_{\mathrm{gas}}$= 753~K, $N$($^{13}$CO)$\simeq$
10$^{17}$ cm$^{-2}$, $N$($^{12}$CO)$\simeq$ 6.7 $\times$ 10$^{18}$
cm$^{-2}$). The results are summarised in Table~\ref{RotDiagTable}.
The LTE level population and the non detection of
high-velocity gas in emission in the millimeter indicate that the warm
component arises in a low surface area, high density region
($n_{\mathrm H}>$10$^{7}$ cm$^{-3}$) with a very small filling factor
in the millimeter.

\subsubsection{High velocity gas}\label{high_velocity_gas}
   
   \begin{figure}  
     \centering
     \resizebox{\hsize}{!}{\includegraphics[]{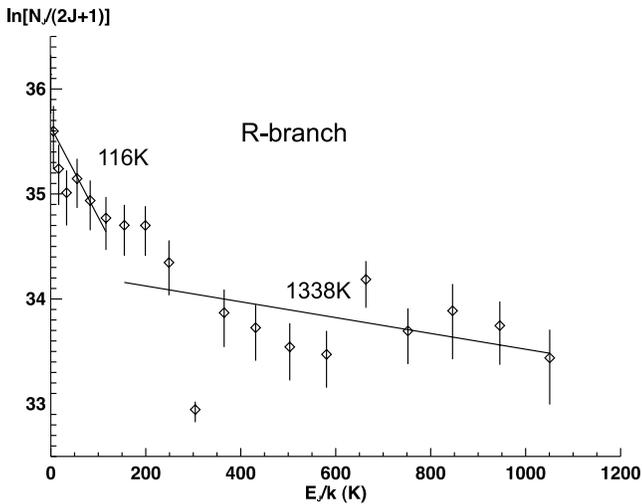}}
   \caption{Rotational diagram of the blue-wing component of $^{12}$CO. \label{fig_diag_wing}}
\end{figure}
The blue absorption component in the {\em ISAAC} data was assumed
optically thin, because the wing is absent in $^{13}$CO, as testified
by the $^{12}$CO and $^{13}$CO profiles in
Fig.~\ref{fig_profile}. We constructed a rotational diagram,
  displayed in Fig.~\ref{fig_diag_wing}, taking into account the
  transitions from levels detected in the $R-$branch and $b_D$=11 km
  s$^{-1}$ to correct for optical thicknesses. Two gas components have
  been fitted to the data. Warm gas at $\sim$116$^{+53}_{-28}$~K and
  with column density of
  2.0$^{+1.7}_{-0.9}$~$\times$~10$^{17}$~cm$^{-2}$ and hot gas at
  $\sim$~1338$^{+2298}_{-518}$~K and column density of
  3.3$^{+5.4}_{-1.2}$~$\times$~10$^{17}$~cm$^{-1}$ provide the best
  fit (Table~\ref{RotDiagTable}). The total column density of
  high-velocity gas is a factor 10 smaller than that of the main
  absorbing gas using $^{12}$CO, assuming that the $^{12}$CO lines are
  optically thin.

\subsubsection{CRIRES components}

\begin{figure*}  
     \centering  
\includegraphics[height=19cm,angle=90]{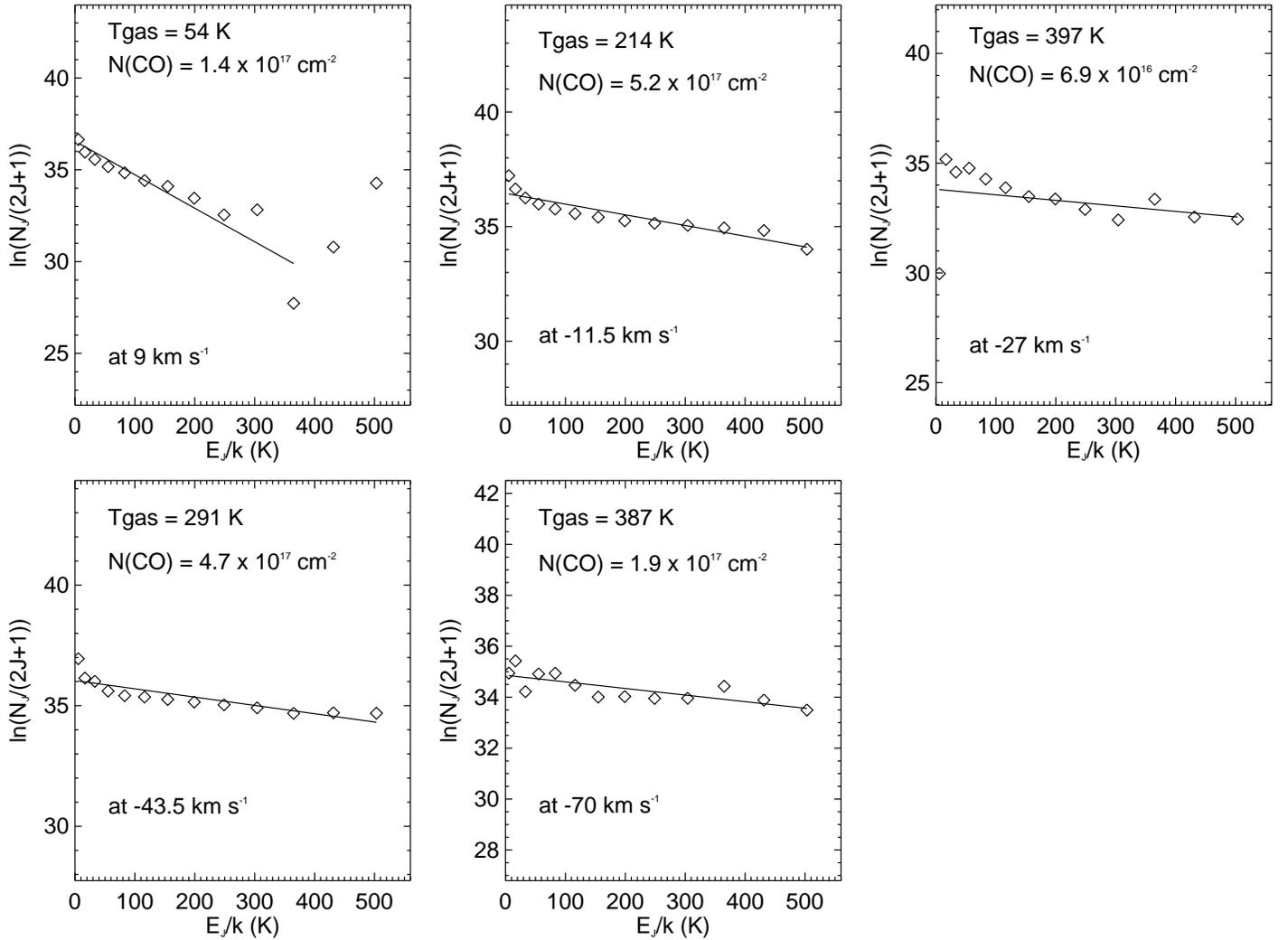}
     \caption{Rotational diagram of $^{12}$CO for each substructure
       components in the {\em CRIRES} data. Equivalent width for $P$(12) and $P$(13) were not used in the analysis of the 9 km s$^{-1}$ substructure because the absorption features are compromised by strong telluric absorption (see Fig.~\ref{fig_muliple_co_crires}). \label{fig_diag_crires}}
\end{figure*}

We drew the rotational diagram for $^{12}$CO for the individual
substructures in Fig.~\ref{fig_diag_crires} and the results are
summarized in the lower part of Table~\ref{RotDiagTable} . The {\em
  CRIRES} lines probe the cool gas towards LLN~19 with substructure
temperatures ranging from 47 to 396 K.

High-$J$ $^{13}$CO transitions up to $R$(22) were available in the
{\em ISAAC} data set to constrain the high-temperature gas component
but they were not observed with {\em CRIRES}. The lack of
high-temperature gas tracers explains the lower overall gas
temperature derived from the {\em CRIRES} data. Optically thick
$^{12}$CO lines were used to derive the column densities with the
optical depth corrections that were applied to the blue-wing {\em
  ISAAC} data ($b_D=11$ km s$^{-1}$). The total column density for the
main component, which comprises the components at +9, -11.5, and -27
km s$^{-1}$, is 7.3 $\times$ 10$^{17}$ cm$^{-2}$, a factor $\sim$10
lower than the value derived from the $^{13}$CO {\em ISAAC} data. The
column density of the component at -27 km s$^{-1}$ is uncertain by a
factor $\sim$25. The column densities derived from {\em CRIRES} data
have large systematic errors because of the assumptions made for the
optical depth corrections (the Doppler width $b_D$ and the number of
absorption features). In the {\em ISAAC} dataset the column density of
the main component is dominated by the column density of warm
gas. Unfortunately the warm gas cannot be probed in the {\em CRIRES}
dataset, which is limited to transitions up to $P$(13) (see
Fig.~\ref{fig_muliple_co_crires}). The systematic errors and the lack
of high-$J$ lines in the {\em CRIRES} dataset may explain the
discrepancy in $^{12}$CO column densities derived from the two
datasets. The two velocity components at -43 and -70~km~s$^{-1}$ in
the {\em CRIRES} data match the {\em ISAAC} blue-wing component well
in terms of temperature ($\sim$~290-390~K) and column density
(1.9--4.7) $\times$ 10$^{17}$ cm$^{-2}$. The reason for the similarity
may stem from the fact that the blue-wing parameters were both derived
from optically thick $^{12}$CO data.

\section{Wind model}\label{wind_model}

In this section we focus on the model-fitting on the {\em
  ISAAC} data since they are more complete in the rotational level
coverage than the {\em CRIRES} data.  We used the velocity structure
seen in the {\em CRIRES} data however, to bound the parameters.

\subsection{Model description and fitting}\label{Model_description}

The temperature and column density of the high-velocity gas suggest
that the absorbing gas may be located in the mixing layer between the
outflow and the envelope. The simple analysis of the {\em ISAAC} data
has assumed an arbitrary distinction between a main component and a
blueshifted one. The {\em CRIRES} data have shown that the outflow
rate of gas is not continuous but may instead proceed by episodes of
strong mass-losses. However, a simple static model like the one we
adopt here cannot handle dynamical effects, although the model can still
give an average value for the mass loss rate.

We fitted the $^{12}$CO and $^{13}$CO synthetic transmission spectrum
of a wind model to the transmission spectrum of LLN~19. The
model focuses on the molecular part of the outflow only.

The spherical wind component of the model is a simplified version of
the parametric model of \citet{Chandler1995ApJ...446..793C}. The wind
is divided into concentric shells located at distances $R$ from the
star. The wind is launched from $R_{in}$ and the size of the spherical
wind is $R_{out}$.  We assume an accelerated wind with velocity of the
form:
\begin{equation}
\displaystyle
V_{w} = \left(V_{max}-V_{R_{in}}\right)\left(1-\frac{R_{in}}{R}\right)^q+V_{R_{in}},
\end{equation}
where $V_{R_{in}}$ is the wind velocity at $R_{in}$ and $V_{max}$ is
the asymptotic wind velocity. The index $q$ controls the variation of
the wind velocity with $R$. The wind velocity does not vary in
  time in our model.

Assuming a constant mass-loss rate $\dot{m}$ (in M$_{\odot}$
yr$^{-1}$) throughout the wind, the spherical wind velocity, mass-loss
rate and column density are related by the mass conservation equation
\begin{equation}
\displaystyle
n(^{12}\mathrm{CO})= \frac{\dot{m}}{4\pi m_{\mathrm{H}_2}R^2 V_{w} (1-\cos{\phi})}\left(\frac{n(\mathrm{^{12}CO})}{n(\mathrm{H}_2)}\right),
\label{eqn_mass_loss_rate}
\end{equation}
\noindent where $m_{\mathrm{H}_2}$ is the gas molecular mass
($m_{\mathrm{H}_2}=2.2$ a.m.u.), $n(\mathrm{^{12}CO})/n(\mathrm{H}_2)$
is the $^{12}$CO abundance relative to H$_2$, and $\phi$ is the
semi-opening angle of the wind (for a spherical wind,
$\phi=\pi/2$). For a spherical wind, a shell of thickness $\Delta
  R$ has a column density $\Delta
  N(^{12}$CO)($R$)=$n(^{12}$CO)($R$)$\times \Delta R$ and
  $\phi=\pi/2$.  We assumed that the wind is fully molecular and that
$n$($^{12}$CO)/$n$(H$_2$)=$10^{-4}$. 

The gas temperature was not computed self-consistently, instead we
adopted a power-law $T_{rot}=T_{in}(R/R_{in})^{-p}$. The gas
temperature power-law index can take positive and negative values. If
the wind cools radiatively and adiabatically by expansion, then $p$ is
positive. The wind can also be heated by shocks and turbulence. In
that case the value for $p$ can be negative.  The $^{12}$CO and
$^{13}$CO level population are at LTE.

Spectra were computed by a simple line+continuum ray-tracing with
$T_{gas}=T_{dust}=T_{rot}=T_{vib}$. The ray-tracing integrates the
radiative transfer equation from shell $i-1$ to shell $i$ for one
  ray taking into account dust \citep{Draine1984ApJ...285...89D} and
CO gas opacities \citep{Chandra1996A&AS..117..557C}:
\begin{equation}
\displaystyle
I_\nu(i) = I_\nu(i-1)e^{-\tau_\nu(i)}+B_\nu\left(T(i)\right)\left(1-e^{-\tau_\nu(i)}\right).
\end{equation}

We removed a low-order polynomial as continuum to the intensity spectra
to obtain synthetic transmission spectra $T_{mod}$.

The number of free parameters amounts to 8: $\theta$=[$R_{in}$,
$R_{out}$, $V_{R_{in}}$, $V_{max}$, $q$, $\dot{m}$, $T_{in}$,
$p$]. The intrinsic turbulent line width $b_D$ is set to
10 km s$^{-1}$ and the $^{12}$CO/$^{13}$CO ratio is assumed to be
65. The number of data points $n$ is 1020 (1024 minus 2 points at each
edge).  The number of degrees of freedom ($dof$) is thus 1012. The
high number of parameters means that degeneracy between parameters is
unavoidable.  We also imposed physically-motivated limits to the
parameters. Assuming an outer radius of 100--500 AU, we can estimate the
mass-loss rate using equation~\ref{eqn_mass_loss_rate}, a CO column
density of 6.7 $\times$ 10$^{18}$ cm$^{-2}$ and a wind velocity of 50
km s$^{-1}$. We found that the mass-loss rate is of the order of
10$^{-7}$ M$_\odot$ yr$^{-1}$. The mass-loss rate ranges from
10$^{-8}$ to 10$^{-6}$ M$_\odot$ yr$^{-1}$ during the fitting
procedure. The wind acceleration parameter $q$ was allowed to vary
between 0 and 3. The gas temperature can decrease (positive $p$
values) or increase (negative $p$ values) as the wind expands.

We used a non-linear global optimisation code based on the Shuffled
Complex Evolution (SCE) algorithm, which mixes evolutionary and
simplex algorithm \citep{Duan1993} , to minimise the Residual Sum
  Squared (RSS) between observational data and model outputs
\begin{equation}
\displaystyle
F(\theta)=RSS=\sum_{j=1}^{n}\left(T_{obs}({\tilde{\nu}}_j)-T_{mod}({\tilde{\nu}}_j,\theta)\right)^2=\sum_{j=1}^{n}f_j^2(\theta),
\end{equation}
where $T_{\mathrm{obs}}(\tilde{\nu}_j)$ is the observed transmission
depth at wavenumber $\tilde{\nu}_j$ and
$T_{\mathrm{mod}}({\tilde{\nu}}_j,\theta)$ is the corresponding
modeled transmission with the $k=11$ parameters. $F(\theta)$ is the
residual sum squared (RSS) or merit function and is related to the
chi-squared. If the noise $\sigma$ is uniform, by definition
$\chi_\nu^{2}$=$\chi^{2}$/dof = (RSS/$\sigma^2$)/dof, where dof is the
degree of freedom. Minimizing the RSS is equivalent to minimizing
$\chi^{2}$. We obtained a reduced $\chi_\nu^{2}$ of 0.4, which
suggests that the noise level is overestimated in average (i.e.,
the S/N of 99 is underestimated).

The SCE is a hybrid algorithm that uses a Genetic algorithm with local
population evolution replaced by a Nelder-Mead Simplex search
\citep{NelderMead1965}. The Genetic algorithm part ensures that a
global minimum can be found while the Nelder-Mead simplex method
provides fast local minimum searches.  The SCE method is efficient in
finding the global minima of non-linear problems with multi local
minima. 

\begin{figure*}  
     \centering
\includegraphics[height=13cm,angle=0]{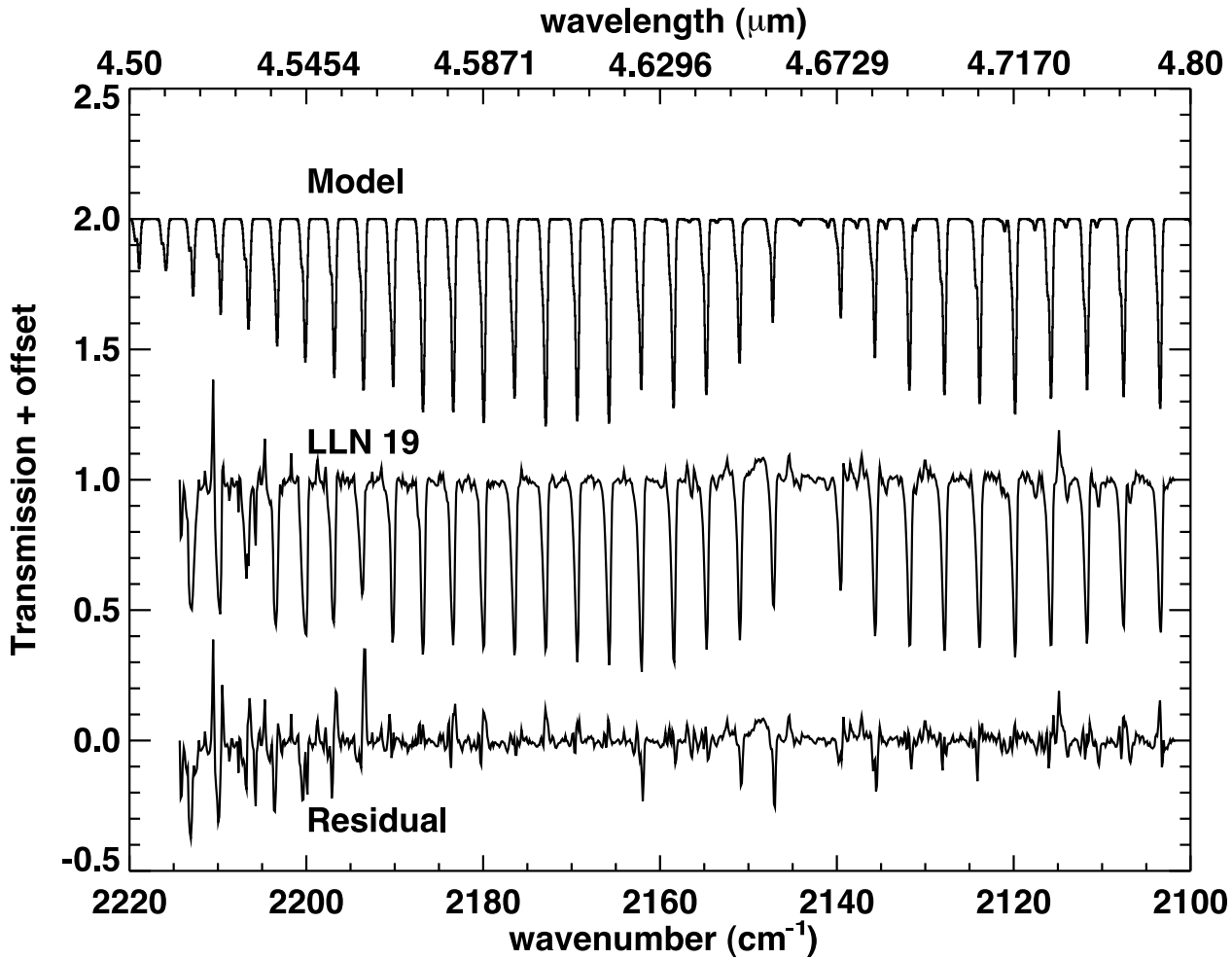}
\caption{Best fit by the wind model. The upper graph shows the best
  model synthetic spectrum shifted by +2.0, the middle graph is the
  observed {\em ISAAC} transmission spectrum shifted by +1.0, and the
  lower graph shows the residual to the best fit. The best fit does
  not completely account for absorptions of very low- and very
  high-$J$ transitions. \label{fig_best_fit}}
\end{figure*}

\subsection{Model results and error budget}\label{Model_parameters}

The best synthetic spectrum is compared to the observations in
Fig.~\ref{fig_best_fit}. After subtraction of the observed
transmission by the wind model, the residual still show weak
absorption in the $J<$~3 lines, testifying of the presence of cold
($T<$~100~K) material, maybe unrelated to LLN~19 where water and CO
ice are present. This cold component was analysed in
Sect.~\ref{cog_analysis}. The fit also fails to explain the relatively
noisy high-$J$ absorption lines above $\tilde{\nu}$= 2200 cm$^{-1}$

For nonlinear models there is no exact way to obtain the
  covariance matrix for the parameters. Instead we use the asymptotic
  approximation and provide the 95\% confidence interval (see Appendix
  for more details). The parameters are relatively well constrained
  (see Table \ref{table_co_wind_env_fit}). The asymptotic error method
  gives unrealistically small errors on the parameters. In addition we
  provide the possible range for the parameters where $\chi_\nu^2<$
  best $\chi_\nu^{2}+1$.

  The best fit parameters for the wind model are summarised in
  Table~\ref{table_co_wind_env_fit}. The wind base is at 1~AU and
  reaches 480~AU. Infrared absorption measurements do not
    constrain the spatial extent of the gas along the line-of-sight.
    Assuming gas densities greater than 10$^6$ cm$^{-3}$ for the CO
    rotational level population to be at LTE and column density of
    5$\times$10$^{22}$ cm$^{-2}$, the outflow radius should be smaller
    than 4500~AU, consistent with our derived value. The total
  $^{12}$CO column density in the wind is 3.5 $\times$ 10$^{18}$
  cm$^{-2}$, lower but comparable to the total of 9.9
  $\times$~10$^{18}$ cm$^{-2}$ found using the simple two slab
  analysis. The wind temperature is 308~K at the base at 1~AU and it
  stays relatively constant through the warm wind with a slight
  increase ($p$=-0.15).  However, the temperature variation is also
  consistent with a temperature decrease within the errors.

The asymptotic maximum wind velocity $V_{max}$ is 87 km s$^{-1}$
starting at 16 km s$^{-1}$. The wind velocity index $q$ is 1.1. The
mass-loss rate is 4.2 $\times$ 10$^{-7}$ M$_\odot$ yr$^{-1}$.

  
\begin{table}  
\centering
\caption[]{Best fit parameters to the {\em ISAAC} data. No velocity correction has been applied. C. I. means confidence interval computed from the asymptotic method. The $\chi_\nu^{2}+1$ Confidence interval is also provided.\label{table_co_wind_env_fit}}
\begin{tabular}[!ht]{llllll}
\noalign{\smallskip}
\hline
\hline
\noalign{\smallskip}
 & Units & Best fit & \multicolumn{1}{c}{95\% C.~I.}&\multicolumn{2}{c}{best $<(\chi_\nu^{2}+1)$ C.~I.} \\
 & & value & \\
\noalign{\smallskip}  
\hline
\noalign{\smallskip}
 $R_{in}$ & AU  &  1.0 & $\pm$ 0.5 & 0.1 & 3.9\\
 $R_{out}$& AU  &  480 & $\pm$ 99 & 74 & 3668\\
 $T_{in}$ & K   & 308 & $\pm$ 10 & 69 & 925 \\
 $p $ &  & -0.15 & $\pm$ 0.26 & -0.32 & 0.40 \\
 $V_{R_{in}}$ & km s$^{-1}$ &  16.1 & $\pm$ 0.03 & 14 & 19\\
 $V_{\mathrm max}$  & km s$^{-1}$ & 87 & $\pm$ 3 & 11 & 270\\ 
 $q$ & & 1.1 & $\pm$ 0.3 & 0.01 & 2.0\\
 $\dot{m}$ & 10$^{-7}$ M$_\odot$ yr$^{-1}$ & 4.2 & $\pm$ 0.7 & 0.3 & 11.6\\
\hline
$S/N$ & & 99\\
$\chi_\nu^2$ & & 0.4 \\
\noalign{\smallskip} 
\hline
\noalign{\smallskip}
\end{tabular}
\end{table}  
%

 
\section{Wind-envelope model}\label{wind_envelope_model}

In this section, we attempt to fit the {\em CRIRES} data with the same
code but with added parameters because the low-$J$ lines indicate that
extra material at low velocity is present in the line-of-sight. In
addition to the wind-model, we modelled simultaneously the absorption
by a quiescent slab component in addition to the wind. The slab of
column density $N_{env}$ is at a single temperature $T_{env}$ and
moves at velocity $V_{env}$. The widths of the slab and wind absorption
lines are 5 and 10 km s$^{-1}$ respectively as suggested by the values
found in Fig.~\ref{fig_fwhm_EW_crires}.

\begin{figure*}
  \centering
  \resizebox{\hsize}{!}{\includegraphics[angle=0]{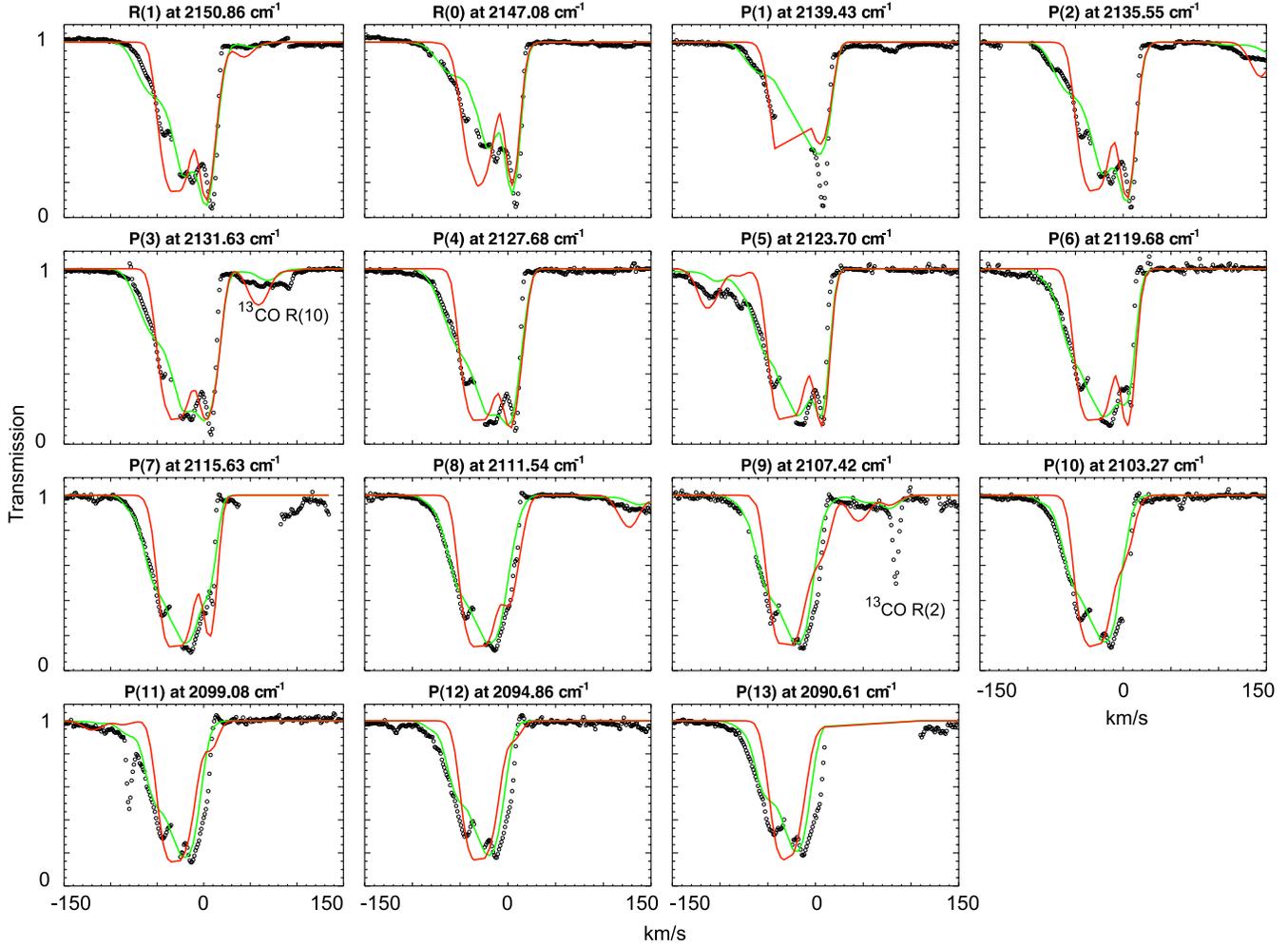}}
   \caption{Two fits to the {\em CRIRES} observations. The red fit results from a wind+envelope model with mass
   loss rate of $\sim$~4.2 $\times$ 10$^{-7}$ yr$^{-1}$ while the green fit results from a model with mass loss rate
   of $\sim$~4.8 $\times$ 10$^{-8}$ yr$^{-1}$. The red fit grossly overestimates the absorption around the -43.5 km s$^{-1}$   features and fails to match the data in the blue tail at $\sim$-70 km s$^{-1}$. On the other hand the green fit under-   estimates the amount of absorption. Both fits stress the failure of the constant mass-loss rate model.
   \label{fig_crires_model_fits}}
\end{figure*}

We performed two series of fits. In the first series the wind
parameters were bracketed around the best values found by fitting the
{\em ISAAC} data. In the second series, all parameters were given the
freedom to take any physically valid value. The best fit in the first
series is plotted in red in Fig.~\ref{fig_crires_model_fits} and the
best fit in the second series is plotted in green. The parameters for
both fits, the $\chi^{2}$ of the fits, and their asymptotic 95\%
Confidence Interval are summarized in
Table~\ref{table_co_wind_env_crires_fit}. We only provide the  
asymptotic errors as the fits are far from optimal ($\chi_\nu^2>100$).
Both fits, which are based on a constant mass-loss rate wind model,
fail to reproduce the complex shape of the $^{12}$CO absorption
profile but the fit with a mass-loss rate an order of magnitude lower
than the value found by fitting the {\em ISAAC} data seems to give a
better match. A closer inspection reveals that both fits struggle in
the region between -30 and -50 km s$^{-1}$. The low mass-loss rate
model underpredicts the amount of absorption whereas the high
mass-loss rate model overpredicts the absorption feature.
The high mass-loss rate fit over-predict the amount of absorption
while the low mass-loss rate fit under-predict the amount of material
in this velocity range. The mass-loss rates provided by the two fits
bracket the mass-loss rates. The temperature of the slab is between 51
and 83 K, warmer than the value derived from the Gaussian fitting
analysis. As expected from the $^{13}$CO {\em ISAAC} data, the
$^{12}$CO column density is three orders of magnitude higher, because
the rotational diagram analysis vastly underestimates column densities
in very optically thick cases. The temperature profile is not well
constrained. Both a slightly increase or decreasing temperature
profile are consistent with the data.

%
  
\begin{table}  
\centering
\caption[]{Best fit parameters to the {\em CRIRES} data for the two series. C. I. means confidence interval. The confidence intervals are approximate statistical errors (see Appendix). Systematic errors are not included.
\label{table_co_wind_env_crires_fit}}
\begin{tabular}[!ht]{llllll}
  \noalign{\smallskip}
  \hline
  \hline
  \noalign{\smallskip}
  & Units & Best fit & \multicolumn{1}{c}{95\% C.~I.}  & Best fit & \multicolumn{1}{c}{95\% C.~I.}\\
  & & low $\dot{m}$ &  & high $\dot{m}$  & \\
  \noalign{\smallskip}
  \hline
  \noalign{\smallskip}
  $R_{in}$ & AU  &  0.14 & $\pm$ 0.02 & 0.9 & $\pm$ 2.6 \\
  $R_{out}$& AU  &  73 & $\pm$ 4 & 167 & -167/+298\\
  $T_{in}$ & K   & 290 & $\pm$ 2 & 340 & $\pm$ 6\\
  $p $ &  & -0.19 & $\pm$ 0.08 & 0.07 & $\pm$ 0.8\\  
  $V_{R_{in}}$ & km s$^{-1}$ &  16.0 & $\pm$ 0.02 & 15.9 & 0.7 \\
  $V_{\mathrm max}$  & km s$^{-1}$ & 70 & $\pm$ 0.6 & 28 & $\pm$ 1\\ 
  $q$ & & 2.2 & $\pm$ 0.08 & 2.3 & $\pm$ 1.2\\
  $\dot{m}$ & 10$^{-7}$ M$_\odot$ yr$^{-1}$ & 0.48 & $\pm$ 0.02 & 4.2 & $\pm$ 3.5\\
  $N_{\mathrm{slab}} $ & 10$^{17}$ cm$^{-2}$ & 4.3 & $\pm$ 0.015 & 6.0 & $\pm$ 1.8\\
  $T_{\mathrm{slab}} $ & K & 51 & $\pm$ 1 & 83 & $\pm$ 3\\
  $V_{\mathrm{slab}} $ & km s$^{-1}$ & -8.9 & $\pm$ 0.03 & -8.6 & $\pm$ 0.02\\
  \hline
  $\chi_\nu^{2}$ & ... & 125 & ... & 224 & ... \\
  \noalign{\smallskip} 
  \hline
  \noalign{\smallskip}
\end{tabular}
\end{table}
%


\section{Discussion}\label{iras08470_discussion}

\subsection{Gaseous and solid CO around LLN~19?}

From our analysis of the gas and solid phase CO, the
$N_{\mathrm{ice}}$(CO)/$N_{\mathrm{gas}}$(CO) ratio is
$\simeq$~5~$\times$~10$^{-3}$. The CO ice/gas is the lowest seen
towards other YSOs in the Vela molecular cloud with similar bolometric
luminosity ($L_*$=1600~$L_{\odot}$ for LLN~19) but close to the value
derived in a few high-mass YSOs ($L_*/L_{\odot}>$10$^{4}$), which have
less than 1\% of CO in the solid state
\citep{Mitchell1990ApJ...363..554M}. Other high-mass YSOs have much
higher fraction of CO in the solid state (i.e., 22.7\% CO ice in the
envelope of RAFGL 7009S \citep{Dartois1998A&A...331..651D}). Also,
contrary to other young stellar objects in the Vela cloud, LLN~19 has
a low water ice abundance with respect to H$_2$ of 1.1 $\times$
10$^{-5}$ and a total water column density of
$N_{\mathrm{ice}}$(H$_2$O)= 1.05 $\pm$ 0.2 $\times$ 10$^{18}$
cm$^{-2}$ \citep{Thi2006}. One explanation is that LLN~19 is in a more
evolved stage than the other YSOs studied in \citet{Thi2006} and has
already cleared its surrounding envelope via powerful outflows. In
this case, both water and CO ices are most likely located in the
foreground material along the line-of-sight toward LLN~19.

Another way to constrain the amount of gas phase CO depletion is to
compare the estimated total column density CO molecules in the
line-of-sight from extinction studies and the detected amount of CO in
the gas phase. The total CO column density in the slab model is 8.9
$\times$ 10$^{18}$ cm$^{-2}$ and 3.5 $\times$ 10$^{18}$ cm$^{-2}$ in
the wind model. In comparison, the fit to the 10 $\mu$m silicate
feature using a 1D radiative transfer code gives $N$(CO)~$\simeq$~8.5
$\times$ 10$^{18}$ cm$^{-2}$ \citep{Thi2006} assuming
[H$_2$]/[CO]=10$^4$, i.e. without CO depletion. The similarity between
the observed and estimated CO column density suggests that CO is
lightly frozen onto grain surfaces along the line-of-sight towards
LLN~19.

We can also estimate the total CO column density from the visual
extinction.  We adopt the conversion factor
N$_{\mathrm{H}}$(CO)/$A_{\mathrm{V}}$ $\sim$ 9 $\times$ 10$^{16}$
cm$^{-2}$ mag$^{-1}$ if there is no CO depletion
\citep{Frerking1982ApJ...262..590F}.  The visual extinction is
estimated from the observed $E(H-K)$ with $A_{\mathrm{V}}$= (15.3
$\pm$ 0.6) $\times$ $(H-K)$ \citep{Rieke1985ApJ...288..618R}.  The
intrinsic $H-K$ color of early B up to F stars is close to zero.
Assuming that the $K$ band flux is dominated by the extinct central
source, $E(H-K)$ $\simeq$ $H-K$. The $H$- and $K$-band magnitudes are
11.92 and 8.88 respectively \citep{Liseau1992A&A...265..577L}, which
translates into an extinction of $\sim$46 magnitude. The derived value
of $N$(CO) is 4.2 $\times$ 10$^{18}$ cm$^{-2}$, smaller but close to
the detected total value of 8.9 $\times$ 10$^{18}$ cm$^{-2}$ in the
2-slab model and very close to the value of 3.5 $\times$ 10$^{18}$
cm$^{-2}$ found using the wind model.  The observed gas-phase CO
column density accounts for the predicted column, indicating that most
CO molecules are in the gas phase.

The two different fitting methods agree on the fact that LLN~19 shows
a much larger amount of warm than cold gas. The temperature of the
cooler component (45~K in the 2-slab model, 200--300~K in the wind
model) is much higher than the sublimation temperature of the pure CO
ice (20~K) and is consistent with the low
$N_{\mathrm{ice}}$(CO)/$N_{\mathrm{gas}}$(CO) ratio of
$\simeq$~~5~$\times$~10$^{-3}$. Because LLN~19 is located in a dense
stellar cluster environment, its outer envelope may be heated above
the CO condensation/sublimation temperature by the ambient radiation
field generated by the nearby stars. This phenomenon has been shown to
work to explain molecular depletion in envelopes around
intermediate-mass YSO \citep{Jorgensen2006}.

\cite{Rettig2006ApJ...646..342G} found higher $N_{\mathrm{gas}}$(CO)
than expected from extinction studies of four YSOs and suggest that
dust grains are settling in the disc midplane. The presence of an
outflow may be considered an indirect evidence for the existence of a
circumstellar disc around LLN~19. The high velocity attained
by the outflowing gas and the absence of redshifted absorption
indicate that the circumstellar disc would be seen at low inclination
with respect to the axis of rotation and the line-of-sight towards
LLN~19 would intercept a small amount of disc material only.
However, the extinction in the wind itself ($A_{\mathrm V}>$50) is high 
enough to mask the redshifted absorption by recessing outflowing gas.

A low CO depletion found in our study concurs with observations of the
CS millimeter line by \citet{Giannini2005A&A...433..941G} which
suggest that no molecular depletion occurs in the Vela Molecular Cloud
on a global scale.

\subsection{$^{12}$CO/$^{13}$CO ratio}

\cite{Goto2003ApJ...598..1038G} compared optically thin $^{12}$CO
$v=2\leftarrow 0$ and $^{13}$CO $v=1\leftarrow 0$ towards three YSOs.
They derived a $^{12}$CO/$^{13}$CO ratio of 137~$\pm$~9, 86~$\pm$~49
and 158 toward LkH$\alpha$~101, AFGL~490, and Mon~R2~IRS3
respectively. \cite{Scoville1983ApJ...275..201S} found a
  $^{12}$CO/$^{13}$CO ratio of $\sim$~100 toward the BN/KL object.
The ratios are 1.5--2 times higher than the generally accepted
interstellar value.  We estimated a $^{12}$CO/$^{13}$CO ratio of 67
$\pm$ 3 towards LLN~19, close to the average interstellar value. The
line-of-sights in Goto et al.\ study probe quiescent ($dv$=2--3.5 km
s$^{-1}$) cold foreground clouds whereas our data suggest that most of
the CO gas is located in a turbulent warm region around LLN~19.

More recently, a detail study of CO isotopologues absorption in the
$M$-band with {\em CRIRES} toward the protoplanetary disc VV CrA and
the YSO Reipurth 50 gives a value of $\zeta(^{12}$C/$^{13}$C) of 100
\citep{Smith2009arXiv0906.1024S}. One possible explanation for the
higher ratio is selective photodissociation of $^{13}$CO in
protoplanetary discs and protostellar environment. It appears that the
$^{12}$CO/$^{13}$CO may be inhomogeneous within a cloud. This supports
the idea that local differentiation processes occur. Detailed analysis
of the $^{12}$CO/$^{13}$CO ratio using optically thin $v=2\leftarrow 0$
$^{12}$CO absorption of the cold component at 9 km s$^{-1}$ will be
published elsewhere (Smith R. et al.).  Interestingly, the
$\zeta(^{12}$C/$^{13}$C) for CO ice has been found to be $\sim$~70
\citep{Boogert2002ApJ...577..271B,Pontoppidan2003A&A...408..981P},
close to the quiescent interstellar gas phase values (60--80).

\subsection{Properties of the outflowing gas}  
We derived some properties of the outflowing gas in
LLN~19 and summarised them in Table~\ref{table_wind_parameters}.

The geometry of absorbing gases cannot be constrained
with spatially unresolved observations. The gas may be outflowing in
form of spherical shells or in highly collimated jets.  For
simplicity, we have assumed that the wind is spherical, purely
molecular, and that the $n$(CO)/$n$(H$_2$) ratio is 10$^{-4}$.

LLN~19 has a luminosity of 1600~L$_{\odot}$. Its inferred mass-loss
rate of 4.8 $\times$ 10$^{-8}$ -- 4.2 $\times$ 10$^{-7}$ M$_\odot$
yr$^{-1}$ is two to three orders of magnitude lower than that in mid-
to early-B ($L_*>10^4~$L$_{\odot}$) YSOs $\sim$~10$^{-5}$ to a few
10$^{-3}$ M$_{\odot}$ yr$^{-1}$ (\citealt{Mitchell1991ApJ...371..342M};
\citealt{Arce2007prpl.conf..245A}). Low mass-loss rates of the order
of 10$^{-7}$ M$_\odot$ yr$^{-1}$ are found for low-mass ($L_*\leq
1$~L$_\odot$) young stars \citep{Bontemps96}. Our result is marginally
consistent with a scaling of the mass-loss rate with luminosity
\citep{Nisini1995A&A...302..169N}.

The source of the momentum carried by extended bipolar outflows
  remains undetermined. By computing the momentum of the young
  outflows we may compare them to values derived from millimeter
  observations of extended outflows. The assumption of spherical flow
  in our modelling implies that the derived momenta are lower
  limits. The momentum per unit area of a shell, assuming a thin shell
  compared to its radius $R$, is
\begin{equation}
p_{\mathrm{shell}}(R)=\dot{m}\frac{dR}{4\pi R^2},
\end{equation}
and the momentum at the outer radius $R_{\mathrm{out}}$ is
\begin{equation}
P = \int_{R_{\mathrm{in}}}^{R_{\mathrm{out}}} \dot{m} dR\simeq \dot{m} R_{\mathrm{out}}.
\end{equation}

The outer radius of the wind from the model is
$R_{\mathrm{out}}\simeq$ 73--480~AU (or (1--7.2) $\times$ 10$^{15}$
cm), which is consistent with values found for high-mass stars.  The
momentum per unit area in the wind at the outer radius is given by
\citep{Mitchell1991ApJ...371..342M}:
\begin{equation}
p(R_{\mathrm{out}})=N(CO)\left(\frac{n(H_2)}{n(CO)}\right)m(H_2)V_w=\frac{\dot{m}}{4\pi R_{\mathrm{out}}},
\end{equation}

We derived a momentum per unit area $p(R_{\mathrm{out}})$ of
(2.9--8.4)~$\times$~10$^{2}$ g cm$^{-1}$ s$^{-1}$ and a total momentum
$P$ of (0.19--9.6)~$\times$~10$^{-4}$ M$_{\odot}$~km~s$^{-1}$.  The
momentum per unit area is $\sim$~100 times and the total momentum is
also $\sim$~1000 times smaller than in outflows around high-mass YSOs
\citep{Mitchell1991ApJ...371..342M}. The momentum rate (also called
momentum flux or force) of (0.4--7.9)~$\times$~10$^{-5}$
M$_{\odot}$~km~s$^{-1}$ yr$^{-1}$ is also similar to the rates found
for outflows from low-mass protostars \citep{Arce2007prpl.conf..245A}.

We also estimated the dynamical timescale of the wind, which corresponds to the time for a gas shell to reach the wind outer radius $R_{\mathrm{out}}$ from $R_{\mathrm{in}}$. The mass in a shell at distance $R$ from the source is:
\begin{equation}
m_{\mathrm{shell}}=dm = 4 \pi R^2 m_{H_2} dN
\end{equation}
The constant mass-loss rate can be rewritten as
\begin{equation}
\dot{m}=\frac{dm}{dt}=4\pi R^2 m_{H_2} \frac{dN(R)}{dt}\simeq 4\pi R_{\mathrm{out}}^2 m_{H_2}\frac{N(R_{\mathrm{out}})}{t_{\mathrm{dyn}}}.
\end{equation}
The dynamical timescale is thus
\begin{equation}
t_{\mathrm{dyn}}\simeq 4 \pi R_{\mathrm{out}}^2 m_{H_2} \frac{N(R_{\mathrm{out}})}{\dot{m}},
\end{equation}

with $N=10^4\times N(^{12}$CO)~=~3.5~$\times$~10$^{22}$ cm$^{-2}$ at
$R_{\mathrm{out}}$. We find that $t_{dyn}\sim$~28 yr at the outer
radius. Using the fits to the {\em CRIRES} data, we estimate the
  dynamical timescale to range between 5 and 28 years. Therefore,
  whatever the actual value, the dynamical timescale is much shorter
  than the values derived from extended molecular outflows in the
  millimeter.

The dynamic timescale is consistent with the values found for
high-mass YSOs (2--200 yr).
The total momentum and the dynamical timescale for the extended
outflow obtained by mapping the environment of LLN~19
(8~$\arcmin$~$\times$~4~$\arcmin$) in $^{12}$CO $J=1 \rightarrow 0$ is
171 M$_{\odot}$~km~s$^{-1}$ and $t_{dyn}\sim$~1.6 $\times$ 10$^5$ yr,
\citep{Wouterloot1999A&AS..140..177W}. Those values were derived
assuming a distance of 2.24 kpc. The discrepancy between the
millimeter and infrared values suggests that the CO fundamental lines
are probing the recent history ($<$~few 100 years) of the outflow.
Outflows change morphology with time \citep{Arce2007prpl.conf..245A}.
The youngest outflows tend to be highly collimated or include a very
collimated component while older sources are less collimated with
wider opening angles. Finally, the sharp drop in total momentum
indicates that the outflow was probably much more active in the past.

\begin{figure}  
     \centering
\includegraphics[height=8cm,angle=0]{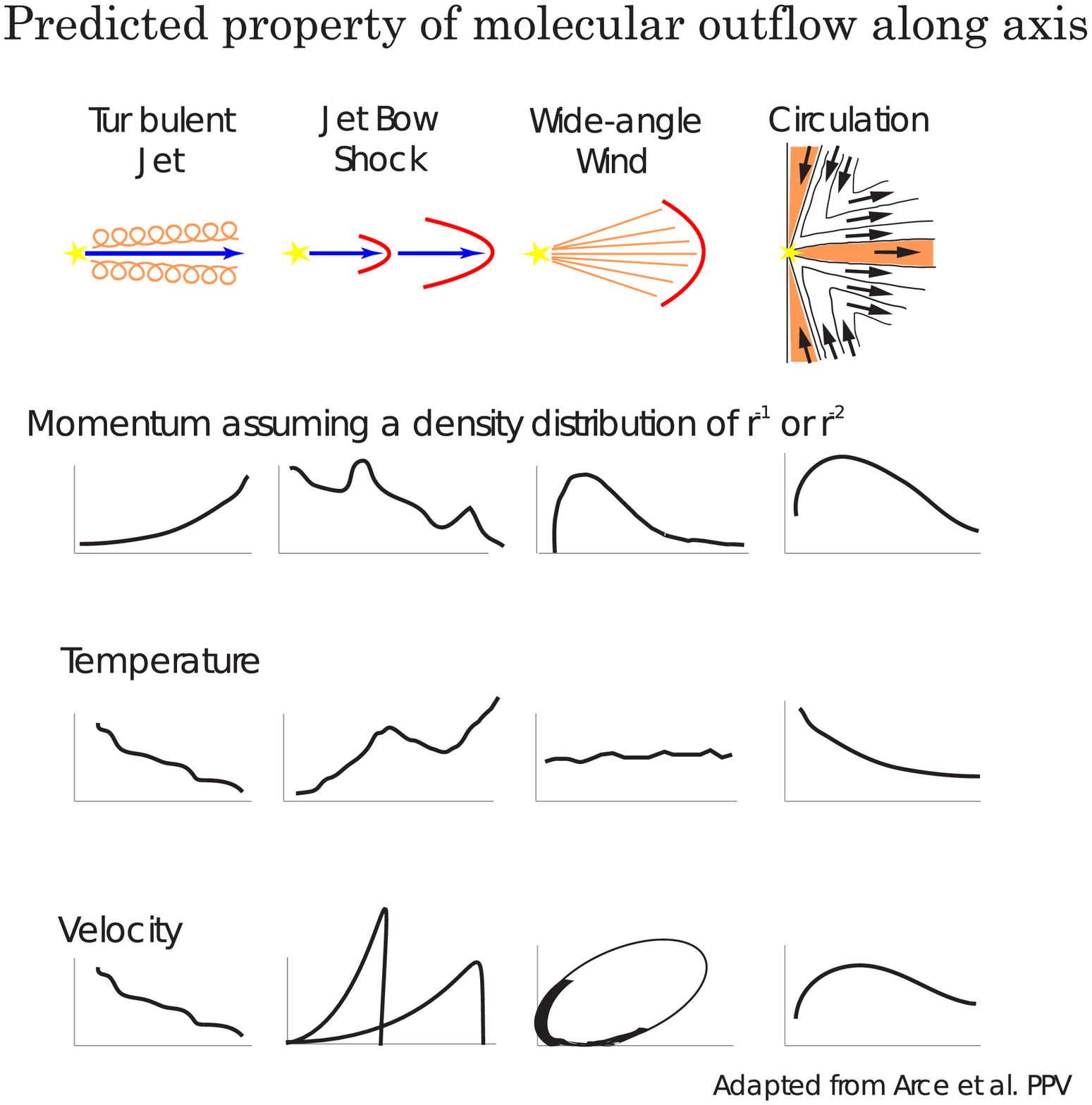}
\caption{Schematics showing different outflow models. The outflow
  momentum, temperature, and velocity profile along the axis are shown
  for the four major types of models: turbulent jet, jet bow shock,
  wide-angle wind, and circulation. The schematics are adapted from
  \citet{Arce2007prpl.conf..245A}. \label{fig_outflow_models}}
\end{figure}

The high-resolution {\em CRIRES} data confirm the picture drawn from
the fitting of the medium-resolution {\em ISAAC} data. The
substructures seen in the CO absorption lines at different velocities
further suggest that the outflow mass-loss is episodic with at least
two major events. We bracketed the mass-loss rate between 0.48
  and 4.2 $\times$ 10$^{-7}$ M$_\odot$ yr$^{-1}$. This is an order of
  magnitude variation in mass-loss and can be achieved in the context
  of FU Orionis type burst
  (e.g., \citealt{Calvet1993ApJ...402..623C}). Multiple blue-shifted
  absorptions are also found in high-mass YSOs and are attributed to
  episodic mass-loss events \citep{Mitchell1991ApJ...371..342M}.

Figure~\ref{fig_outflow_models} summarizes schematically the four
major outflow models.  From all the four models, only the jet bow
shock model predicts episodic variation in jet properties as observed
towards LLN~19 \citep{Arce2007prpl.conf..245A}. Also the jet bow shock
model is the sole model that can explain a temperature and velocity
increase with distance from the outflow source.

\begin{table*}  
\centering
\caption[]{Outflow parameters. \label{table_wind_parameters}}
\begin{tabular}[!ht]{lllll}
\noalign{\smallskip}
\hline
\hline
\noalign{\smallskip}
 & Units & {\em ISAAC} & {\em CRIRES} & {\em CRIRES} \\
 &       &             & low  $\dot{m}$ & high  $\dot{m}$\\
\noalign{\smallskip}
\hline
\noalign{\smallskip}
 Total column density $N$(H$_2$) & 10$^{22}$ cm$^{-2}$ & 3.5 & 2.2 & 2.6 \\
Mass loss rate $\dot{m}$ & 10$^{-7}$ M$_\odot$ yr$^{-1}$  & 4.2  & 0.48 & 4.2  \\
Momentum per unit area $p$ & 10$^{2}$ g cm$^{-1}$ s$^{-1}$ & 2.9 & 2.5 & 8.4 \\
Total momentum $P$ & M$_{\odot}$~km~s$^{-1}$ & 9.6~$\times$~10$^{-4}$ & 1.9 $\times$ 10$^{-5}$ & 3.3 $\times$ 10$^{-4}$ \\
Momentum rate & M$_{\odot}$~km~s$^{-1}$ yr$^{-1}$ & 7.9~$\times$~10$^{-5}$ &3.9 $\times$ 10$^{-6}$ & 1.2 $\times$ 10$^{-5}$ \\
Dynamical timescale $t_{\mathrm dyn}$ & yr & 26 & 5 & 28 \\
\noalign{\smallskip} 
\hline
\noalign{\smallskip}
\end{tabular}
\end{table*}
Another peculiar feature of the CO lines towards LLN~19 is the
intrinsic line widths.  Generally, $^{12}$CO ro-vibrational
absorptions/emissions toward YSOs are dominated by a hot component at
$T_{gas}>$~600~K (e.g., \citealt{Mitchell1990ApJ...363..554M}).  The
intrinsic width is low 3-5 km s$^{-1}$ in contrast to our result
($dv$=10-12 km s$^{-1}$) for the outflowing gas, although widths up to
8--10 km s$^{-1}$ have been observed. Narrow cold gas found in the
line-of-sight of high-mass YSOs is attributed to gas in the extended
envelope. The wind model-envelope fit to the {\em CRIRES} data also
suggests that a cold envelope may be present in the line-of-sight.

One possibility for the warm gas to show large intrinsic width is
that turbulence is at play. The detection of turbulent gas would be
tantalising. The decay of turbulent eddies could provide the necessary
heating to keep large amount of gas at $\sim$100-1000~K as the
radiative heating from the central star decreases with distances. One
possible origin for the turbulence are Kelvin-Helmholtz instabilities
caused by the interaction of atomic jet entraining molecular material
\citep{Delamater2000ApJ...530..923G}.

Conversely the jet bow shock model of outflows provides a scenario to
generate turbulent gases (see fig.~\ref{fig_outflow_models}) . The
atomic jets from the inner disc surfaces impact the ambient gas
creating shocks. In turn, the high pressure gas between the shocks is
ejected sideways. This high pressure gas interacts with unperturbed
molecular gas, generating turbulences. Shocks will heat the gas to
several hundred Kelvin, consistent with the gas temperatures of the
different substructures. The jet bow model can explain the slight
increase of gas temperature with radius contrary to the turbulent jet
model. However the 95\% confidence interval for the gas temperature
allows decreasing gas temperature with distance to the star (Table
\ref{table_co_wind_env_fit}). A flat temperature profile would be
consistent with the wide-angle wind model. The large uncertainties in
the model parameters, especially for the gas temperature profile,
prevent us to favour a specific outflow model.

\section{Conclusions}\label{iras08470_conclusion}

We analysed medium- ($R\sim$~10,000) and high-resolution
($R\sim$~100,000) $\mathrm 4.5-4.8~\mu m$ spectra of the embedded
intermediate-mass young stellar object LLN~19 in the Vela Molecular
Cloud. Gas phase $^{12}$CO and $^{13}$CO ro-vibrational lines have
been analysed with a curve-of-growth and a rotational diagram, and
also fitted by a wind model. Both analyses give similar values for the
total column density of warm CO gas and a $^{12}$C/$^{13}$C isotopic
ratio of 67~$\pm$~3, lower than values close to 100 found in other
studies using infrared absorption lines. The discrepancy in
$^{12}$C/$^{13}$C isotopic ratios derived from CO may just reflect the
occurrence of varied chemical phenomena in molecular clouds and
star-forming regions.

The CO gas towards LLN~19 arises from an outflowing warm
  gas. Most CO molecules are in the gas phase with less than 0.5\%
  condensed on grain surfaces, maybe in a foreground cloud. The wind
  parameters of LLN~19 are closer to those from low-mass protostars
  than high-mass protostars, as seen in infrared absorption. The warm
  outflowing gas has been created only recently with a dynamical
  timescale of less than 28 yr, if the data are analysed in the context
  of a static spherical wide-angle wind model. However, the kinematics
  of the outflow as seen in the high spectral resolution {\em CRIRES}
  data are more consistent with a episodic jet bow shock model where
  the internal bow shocks lead to high pressure and high temperature
  shells. A dynamical mass-loss rate model is warranted.

LLN~19 (IRAS~08470--4321) is a good example of an object in the phase
of cleaning its envelope by an outflow. Temporal monitoring of LLN~19
with {\em CRIRES} would provide further clues on the evolution of its
outflow.

\section{Acknowledgments}
WFT is supported by a Scottish Universities Physics Alliance (SUPA)
fellowship in Astrobiology. This research is supported by the
Netherlands Research School for Astronomy (NOVA) and a NWO Spinoza
grant. K.M.P. acknowledges a Hubble fellowship. The authors wish to
thank the VLT staff for all their help in obtaining the
observations. WFT thanks Peder Noberg for discussions on statistics
and data fitting.  We thank the referee for his/her useful comments
and suggestions.

\section{Appendix}\label{appendix}

For nonlinear models there is no exact way to obtain the covariance
matrix for the parameters. Nor do the optimization codes give directly
error parameters. Parameter confidence intervals are approximated from
the Hessian matrix of $\tau_{mod}(\theta)$ at the solution
$\theta=\hat{\theta}$ \citep{Donaldson1987,Wall2003psa..book.....W}:
\begin{equation}
H(\hat{\theta})=2J(\hat{\theta})^TJ(\hat{\theta})+2\sum_{j=1}^{n} f_j(\hat{\theta})H_j(\hat{\theta}),
\end{equation} 
where $J(\hat{\theta})$ is the Jacobian of $\tau_{mod}(\hat{\theta})$.
In the neighbourhood of the solution,
$||f_j(\hat{\theta})H_j(\hat{\theta})||$ is neglectable compared to
$||J(\hat{\theta})^TJ(\hat{\theta})||$. Therefore, the Hessian matrix
at the minimum for $F(\theta)$ (equal to the residual sum square RSS)
at $\theta=\hat{\theta}$ can be approximated by
\begin{equation}
H(\hat{\theta})\simeq2J(\hat{\theta})^TJ(\hat{\theta})
\end{equation}
An unbiased estimate of the parameter covariance matrix is 
\begin{equation}
C(\hat{\theta})=2\frac{F(\hat{\theta})}{dof}H(\hat{\theta})^{-1}
=\frac{F(\hat{\theta})}{dof}(J(\hat{\theta})^TJ(\hat{\theta}))^{-1},
\end{equation}
where $s^2=F(\hat{\theta})/dof$ is an unbiased approximation of the residual variance $\sigma^2$ ($\sigma\sim s$). The variance of the $i$th parameter $var(\hat{\theta}_i)$ is the $i$th diagonal element of the matrix $C(\hat{\theta})$.
If $\theta_i^*$ is the ''true'' solution, then the $100(1-\phi)$ percentage confidence interval on $\hat{\theta}_i$ is
\begin{equation}
\hat{\theta}_i-\sqrt{var(\hat{\theta}_i)}t^{1-\phi/2}_{dof}<\theta_i^*<\hat{\theta}_i+\sqrt{var(\hat{\theta}_i)}t^{1-\phi/2}_{dof},
\end{equation}
where $t^{1-\phi/2}_{dof}$ is the $100(1-\phi/2)$ percentage point of
the two-tail student $t$-distribution with $dof$ degrees of freedom.
When the number of degrees of freedom exceeds 150, the Student
distribution equals the normal distribution. In our case, a 95\%
confidence interval would be the parameter $\pm 1.96 \times
\sqrt{var(\hat{\theta}_i)}$. The Jacobian matrix element
$J_{ij}(\hat{\theta})$ is approximated by forward finite difference
around $\hat{\theta}$
\begin{equation}
J_{ij}(\hat{\theta})\simeq\frac{\tau_{\mathrm{mod}}({\tilde{\nu}}_j,\theta_i)-\tau_{\mathrm{mod}}({\tilde{\nu}}_j,\hat{\theta_i})}{\theta_i-\hat{\theta_i}}.
\end{equation}
In this paper, the parameter intervals are computed for 95\% confidence.

\bibliographystyle{mn2e}
\bibliography{vela_gas}  
\bsp

\label{lastpage}

\end{document}